\documentclass[letterpaper]{article} 
\usepackage[preprint]{aaai2027}  
\usepackage[hyphens]{url}  
\usepackage{graphicx} 
\usepackage{natbib}  
\usepackage{caption} 
\usepackage{algorithm}
\usepackage{newfloat}
\usepackage{pifont} 
\usepackage{listings}
\DeclareCaptionStyle{ruled}{labelfont=normalfont,labelsep=colon,strut=off} 
\floatstyle{ruled}
\newfloat{listing}{tb}{lst}{}
\floatname{listing}{Listing}

\usepackage{booktabs}

\usepackage[utf8]{inputenc} 
\newcommand{\href}[2]{#2}     
\usepackage{booktabs}       
\usepackage{amsfonts}       
\usepackage{nicefrac}       
\usepackage{microtype}      
\usepackage{graphicx}
\usepackage{amssymb} 
\usepackage{amsmath}
\usepackage{booktabs}
\usepackage{array}
\usepackage{algorithm}
\usepackage{caption}
\usepackage{algpseudocode} 
\usepackage{tabularx}
\usepackage{tcolorbox}
\usepackage{booktabs}
\usepackage{multirow}
\newcommand{\best}[1]{\textbf{#1}}

\newcommand{\logoblog}{\raisebox{-0.2ex}{\includegraphics[height=1em]{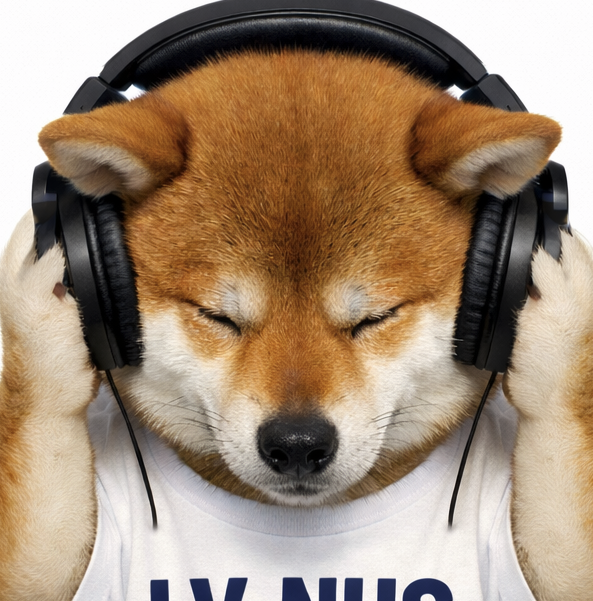}}}
\newcommand{\logohf}{\raisebox{-0.2ex}{\includegraphics[height=1em]{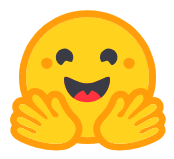}}}
\definecolor{deepblue}{RGB}{0,45,114}

\makeatletter
\def\@maketitle{%
  \vbox{%
    \let\footnote\thanks\relax%
    \hsize\textwidth%
    \linewidth\hsize%
    \vskip 0.625in minus 0.125in%
    \centering%
    {\LARGE\bf \@title \par}%
    \vskip 0.1in minus 0.05in%
    {\Large\textbf{\@author\ifhmode\\\fi}}%
    \vskip .2em%
    {\normalsize \aaai@affiliations\ifhmode\\\fi}%
    \vskip 1.15em%
  }%
}
\makeatother

\title{Audio Interaction Model}
\author{
    Zhifei Xie\textsuperscript{\rm 1}\thanks{\protect\raggedright Equal contribution. Work done during Zihang Liu (liuzihang99@gmail.com)'s visit at NUS-LV. \textsuperscript{\textdagger}Corresponding authors.},
    Zihang Liu\textsuperscript{\rm 2*},
    Ze An\textsuperscript{\rm 2},
    Xiaobin Hu\textsuperscript{\rm 2},
    Yue Liao\textsuperscript{\rm 2},
    \\[0.25em]
    Ziyang Ma\textsuperscript{\rm 1},
    Dongchao Yang\textsuperscript{\rm 3},
    Mingbao Lin\textsuperscript{\rm 2\textdagger},
    Deheng Ye\textsuperscript{\rm 1\textdagger},
    \\[0.25em]
    Shuicheng Yan\textsuperscript{\rm 2\textdagger},
    Chunyan Miao\textsuperscript{\rm 1\textdagger}
}
\affiliations{
    \textsuperscript{\rm 1}NTU \quad
    \textsuperscript{\rm 2}NUS \quad
    \textsuperscript{\rm 3}CUHK\\[0.25em]
    \texttt{Zhifei001@e.ntu.edu.sg}\\[0.9em]
    \logoblog~\textbf{Project page:}
    \href{https://xzf-thu.github.io/Audio-Interaction}{\texttt{\textcolor{deepblue}{https://xzf-thu.github.io/Audio-Interaction}}}\\[0.3em]
    \logohf~\textbf{Data:}
    \href{https://huggingface.co/datasets/zhifeixie/StreamAudio-2M}{\texttt{\textcolor{deepblue}{huggingface.co/datasets/zhifeixie/StreamAudio-2M}}}
}

\begin{document}

\maketitle

\begin{abstract}
Audio is continuous and interactive, yet most Large Audio Language Models (LALMs) remain offline and streaming systems usually specialize in ASR or spoken dialogue.
We formalize the \textbf{\textsc{Audio Interaction Model}}, an always-on \emph{perceive--decide--respond} paradigm that tracks context, decides whether intervention is warranted, and responds without stopping listening.
We instantiate it with \textbf{\textsc{Audio-Interaction}} and introduce \textbf{\textsc{SoundFlow}}, coupling streaming-native data construction, comprehension-aware silence/response supervision, dual-loss training, and asynchronous FIFO inference.
We also construct \textbf{\textsc{StreamAudio-2M}}, a 2.6M-item, 302k-hour corpus spanning 7 capability families and 28 sub-tasks, together with \textbf{\textsc{Proactive-Sound-Bench}}.
Across 8 benchmarks, \textsc{Audio-Interaction} remains competitive on mainstream audio tasks while enabling spoken-instruction robustness, long-stream interaction, and proactive intervention.
\end{abstract}

\begin{figure*}[!t]
\centering
\includegraphics[width=0.9\linewidth]{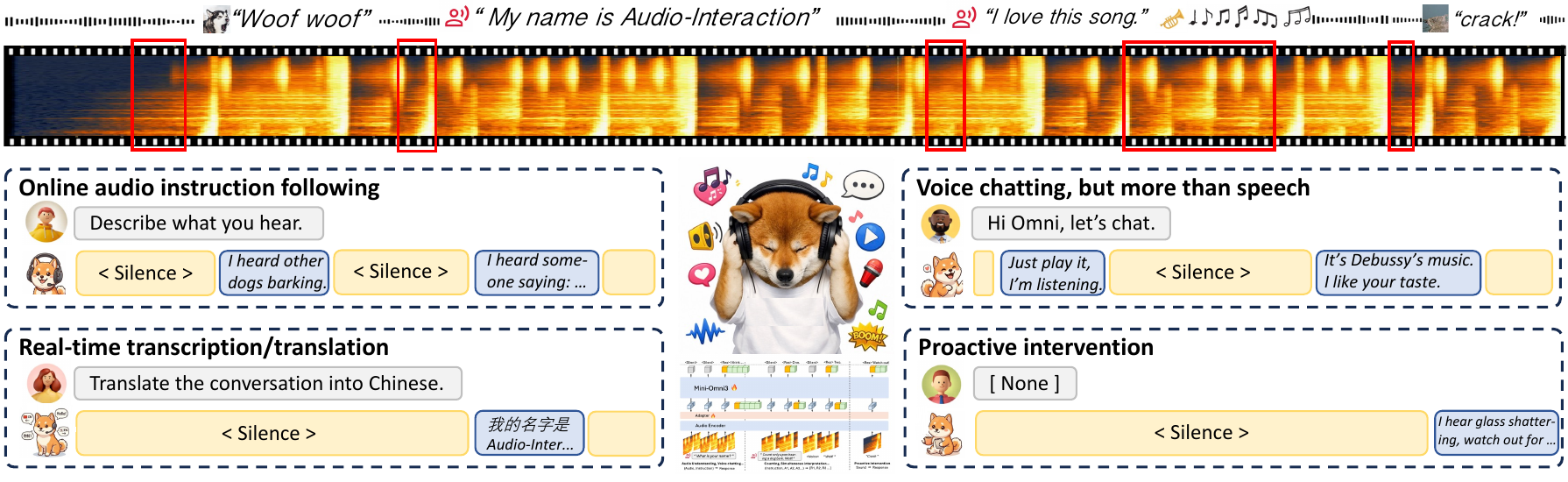}
\caption{\textsc{Audio-Interaction} processes streaming audio and decides at each step whether to stay silent or respond. It unifies offline audio tasks, such as dialogue and ASR, with streaming-native interaction.}
\label{figure1-overview}
\end{figure*}

\section{Introduction}

Audio is a continuous channel through which people perceive events, infer intent, and decide when to act. A useful system should therefore do more than wait for a recording to end. It should listen as sound unfolds, retain context across time, and decide whether the current moment calls for a response. Large audio language models (LALMs) have made rapid progress toward general audio understanding~\citep{chu2023qwen, tang2023salmonn, xie2024mini}, including speech recognition, audio question answering, emotion understanding, and reasoning over acoustic scenes. Yet most of these models inherit the offline interface of multimodal LLMs: given a complete audio clip and an instruction, produce one output.

This interface is poorly matched to real acoustic interaction. In daily use, a model may need to transcribe speech while it is still being spoken, answer a follow-up question after several turns, ignore irrelevant background sounds, or proactively warn about a dangerous event before the user asks. Existing streaming systems address pieces of this problem, but they remain fragmented. Streaming ASR models focus on transcription; spoken-dialogue systems focus on conversational speech; full-duplex models reduce turn-taking latency but usually treat non-speech events as background. Thus, the field faces a gap between \emph{general audio understanding}, where offline LALMs are strong, and \emph{real-time audio interaction}, where current systems remain narrow.

As shown in Figure\,\ref{figure1-overview}, we close this gap by formalizing the \textbf{\textsc{Audio Interaction Model}}. Instead of mapping a finished clip to a single answer, it performs an always-on \emph{perceive--decide--respond} loop. \emph{Perceive} continuously updates the acoustic and interaction context as new audio arrives; \emph{decide} predicts whether accumulated context warrants silence or intervention; and \emph{respond} generates content when triggered, without stopping subsequent perception. This unifies response timing and response content. Conventional tasks such as ASR, speech translation, audio understanding, and spoken dialogue become special cases, while simultaneous interpretation, audio instruction following, long-form interaction, and proactive assistance become natural extensions.

Realizing this loop introduces two coupled challenges absent from offline LALMs. \textbf{(C1) Comprehension-grounded response triggering.} The \emph{decide} stage must infer \emph{when} to respond from accumulated meaning rather than shallow acoustic cues. A cough, a siren, or a hesitant pause may or may not warrant intervention depending on the surrounding stream, but existing datasets rarely provide continuous audio with properly timed silence/response supervision. \textbf{(C2) Context continuity under chunked interaction.} The \emph{perceive} stage must reconstruct acoustic continuity and retain long-range history from fixed-length chunks, while the \emph{respond} stage must not block incoming audio during autoregressive decoding. They are thus dependent: C2 supplies the continuous evidence on which C1 must base its decision, and every response must preserve that evidence for the next decision.

We instantiate the formulation as \textbf{\textsc{Audio-Interaction}}, trained and deployed through \textbf{\textsc{SoundFlow}}. For \emph{perceive}, SoundFlow composes coherent long-form streams, removes stitching artifacts with time-frequency joint preprocessing (\textsc{TFJP}), and uses history-review training to recover and retain cross-chunk context, directly addressing C2 while preventing artifact-based shortcuts. For \emph{decide}, explicit silence/response labels and comprehension-aware negative examples teach when to speak and when to hold back, directly addressing C1 using the context learned above. For \emph{respond}, a dual loss preserves language generation while learning the control decision, and asynchronous FIFO scheduling decouples audio encoding from text decoding so generation does not interrupt perception, completing the solution to C2.

To support this training and evaluation pipeline, we construct \textbf{\textsc{StreamAudio-2M}}, a 2.6M-item, 302k-hour corpus covering 7 capability families and 28 streaming sub-tasks. Each example is a 3--15 turn interaction with heterogeneous acoustic events and sparse, context-dependent response cues, making it better aligned with always-on interaction than conventional clip-level instruction data. We also introduce \textbf{\textsc{Proactive-Sound-Bench}}, a benchmark of 644 human-designed acoustic events that evaluates whether a model can proactively respond to safety-critical or assistance-worthy sounds while abstaining on benign distractors.


Across 8 benchmarks, \textsc{Audio-Interaction} remains competitive with offline LALMs on audio understanding, spoken dialogue, ASR, and speech translation, while enabling streaming-native capabilities such as spoken-instruction robustness, proactive response, and stable long-stream interaction. Further analyses reveal cross-chunk continuity in early decoder layers and shared attention-based silence/response control, suggesting a unified interaction mechanism rather than task-specific triggers. Our contributions are threefold:
\begin{itemize}
    \item We formalize the \textbf{\textsc{Audio Interaction Model}} as a \emph{perceive--decide--respond} paradigm and identify two coupled challenges: comprehension-grounded decisions (C1) and continuous context under chunked interaction (C2).
    \item We introduce \textbf{\textsc{Audio-Interaction}} and \textbf{\textsc{SoundFlow}}, which address C2 through continuous perception and non-blocking inference, and C1 through comprehension-aware silence/response supervision.
    \item We construct \textbf{\textsc{StreamAudio-2M}} and \textbf{\textsc{Proactive-Sound-Bench}}, and validate that the resulting model preserves conventional audio ability while enabling proactive and long-stream interaction.
\end{itemize}

\section{Related Work}
\label{sec:related_work}

\paragraph{Large Audio Language Models.}
Large audio language models typically combine an audio encoder, an adapter, and a language model backbone~\citep{chu2024qwen2, tang2023salmonn, qwen25omni2025}, a designed shared by our Qwen2.5-Omni-based implementation~\citep{qwen25omni2025}.
Recent systems have expanded from speech-centric tasks to broader audio understanding and reasoning~\citep{goel2025audio, xu2025fireredasr}.
But, they assume an offline input-output interface: the audio clip is observed before generation. This makes them powerful perceptual models but weak interaction models.

\paragraph{Streaming Multi-modal Systems.}
Streaming audio models reduce latency by consuming audio incrementally, but they usually specialize in one behavior.
Streaming ASR focuses on transcription~\citep{gao2022paramformer}; spoken-dialogue and full-duplex systems focus on conversational speech~\citep{xie2024mini, fang2024llama, fang2025llamaomni2, defossez2024moshi, qwen25omni2025}
Online video and proactive agent systems explore real-time decision making in other modalities~\citep{li2025videochat, proactive, pask,needllm}, but rely on visual frames or cascaded textualized observations.
%

\begin{figure*}
    \centering
    \includegraphics[width=0.9\linewidth]{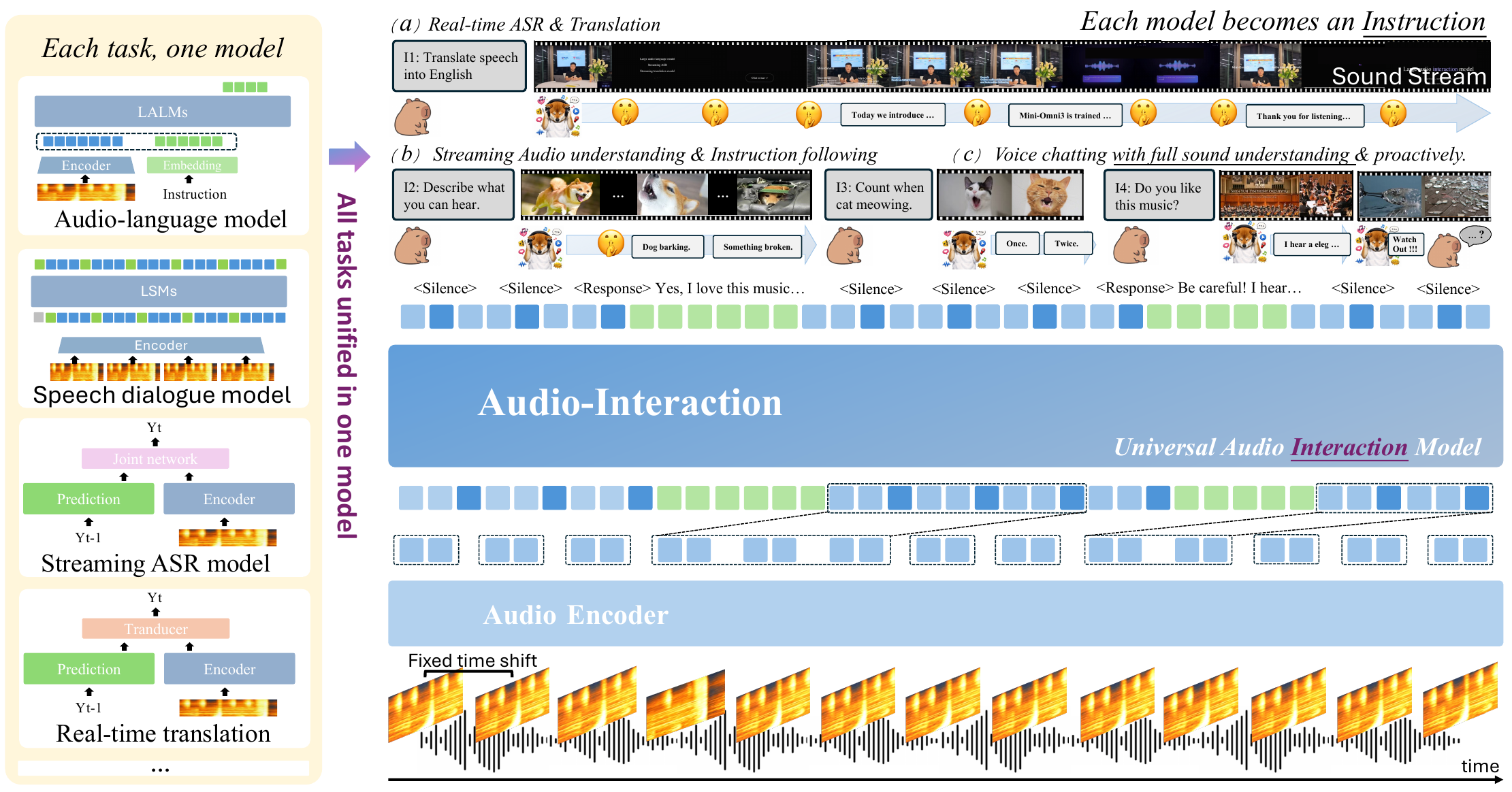}
    \caption{Comparison of offline LALMs, task-specific streaming systems, and \textsc{Audio-Interaction}, which listens continuously and decides whether to stay silent or respond.}
    \label{fig:main}
\end{figure*}

\section{Audio-Interaction}

\subsection{Overview}

As illustrated in Figure\,\ref{fig:main}, conventional LALMs map a completed utterance $\mathcal{A}$ and instruction $x$ to one output, $y=f(x,\mathcal{A})$. In contrast, \textsc{Audio-Interaction} repeatedly executes \emph{perceive--decide--respond} over an incoming stream:
\begin{equation}
\label{eq:pdr_loop}
\begin{aligned}
C_t &= f_{\mathrm{per}}(C_{t-1},a_t),\\
d_t &= f_{\mathrm{dec}}(C_t),\\
r_t &=
\begin{cases}
\varnothing, & d_t=\texttt{<silent>},\\
f_{\mathrm{resp}}(C_t), & d_t=\texttt{<response>}.
\end{cases}
\end{aligned}
\end{equation}

Here $a_t$ is the current audio chunk, $C_t$ is the continuously updated acoustic and interaction context, $d_t$ is the silence/response decision, and $r_t$ is the generated response. This decomposition makes the challenge relationship precise. C2 concerns whether $C_t$ remains coherent and available across chunk boundaries and during response generation; C1 concerns whether $d_t$ is grounded in the meaning represented by $C_t$. Therefore, C2 spans the \emph{perceive} and \emph{respond} stages and is a prerequisite for the C1 decision at the \emph{decide} stage.

\textsc{SoundFlow} implements Equation\,\ref{eq:pdr_loop} through three matched components. For \emph{perceive}, streaming-native data construction and history-aware chunk training establish $C_t$ and address C2. For \emph{decide}, comprehension-aware silence/response supervision learns $d_t$ from that context and addresses C1. For \emph{respond}, dual-loss training preserves response quality, while asynchronous FIFO inference keeps updating the context during decoding and completes the solution to C2.

\subsection{Streaming Data Construction}
\label{sec3.2: Streaming Data Construction}

A model cannot learn the loop from conventional short (audio, instruction, response) triplets. The \emph{perceive} stage requires coherent long-form context to address C2, while the \emph{decide} stage requires sparse, context-dependent silence/response targets to address C1. We therefore construct streams in which events are temporally ordered, background sounds persist across turns, and only a small subset of moments require a response. The construction has two parts.

\textbf{Time-frequency joint preprocessing module.} Because our streams are composed from shorter clips, unnatural boundaries can become spurious trigger cues. We apply a lightweight time-frequency joint preprocessing (\textsc{TFJP}) module before composition. TFJP regularizes temporal gaps and spectral continuity by clipping excessive internal silence (\texttt{silence\_cut}), estimating and removing background noise from low-energy regions (\texttt{noise\_profile} $\rightarrow$ \texttt{denoise}), locating the densest informative span (\texttt{core\_locate}), and refining both boundaries with half-chunk alignment $\delta=\frac{1}{2}$ and short-window spectral smoothing $\omega$ (\texttt{boundary\_norm} $\rightarrow$ \texttt{spec\_smooth}). The early loop stabilizes silence and noise statistics; if the final silence check still changes the segment, the process returns to Stage~1. TFJP preserves acoustic continuity for C2 and prevents boundary artifacts from becoming shortcuts for the C1 decision. The full procedure is shown in the supplementary Algorithm\,\ref{alg:tfjp}.

\textbf{Hierarchical audio event selection.} Smooth boundaries alone do not guarantee meaningful streams. Random concatenation can produce implausible scenes or contradictory events. We therefore use a hierarchical event curation pipeline. \emph{(i) Scenario planning}: an LLM plans a high-level scenario from randomly matched audio annotations, including multiple topics or sub-events. \emph{(ii) Event refinement}: each topic is refined into concrete audio events with target clip specifications. \emph{(iii) Clip grounding}: final clips are obtained by retrieval or generation; retrieval selects and verifies the top-3 candidates from an audio database, while generation synthesizes an event when retrieval fails. This hierarchy yields coherent histories for C2 and makes the C1 trigger depend on accumulated semantics rather than isolated events.

\begin{figure}[!t]
\centering
\includegraphics[width=0.92\linewidth]{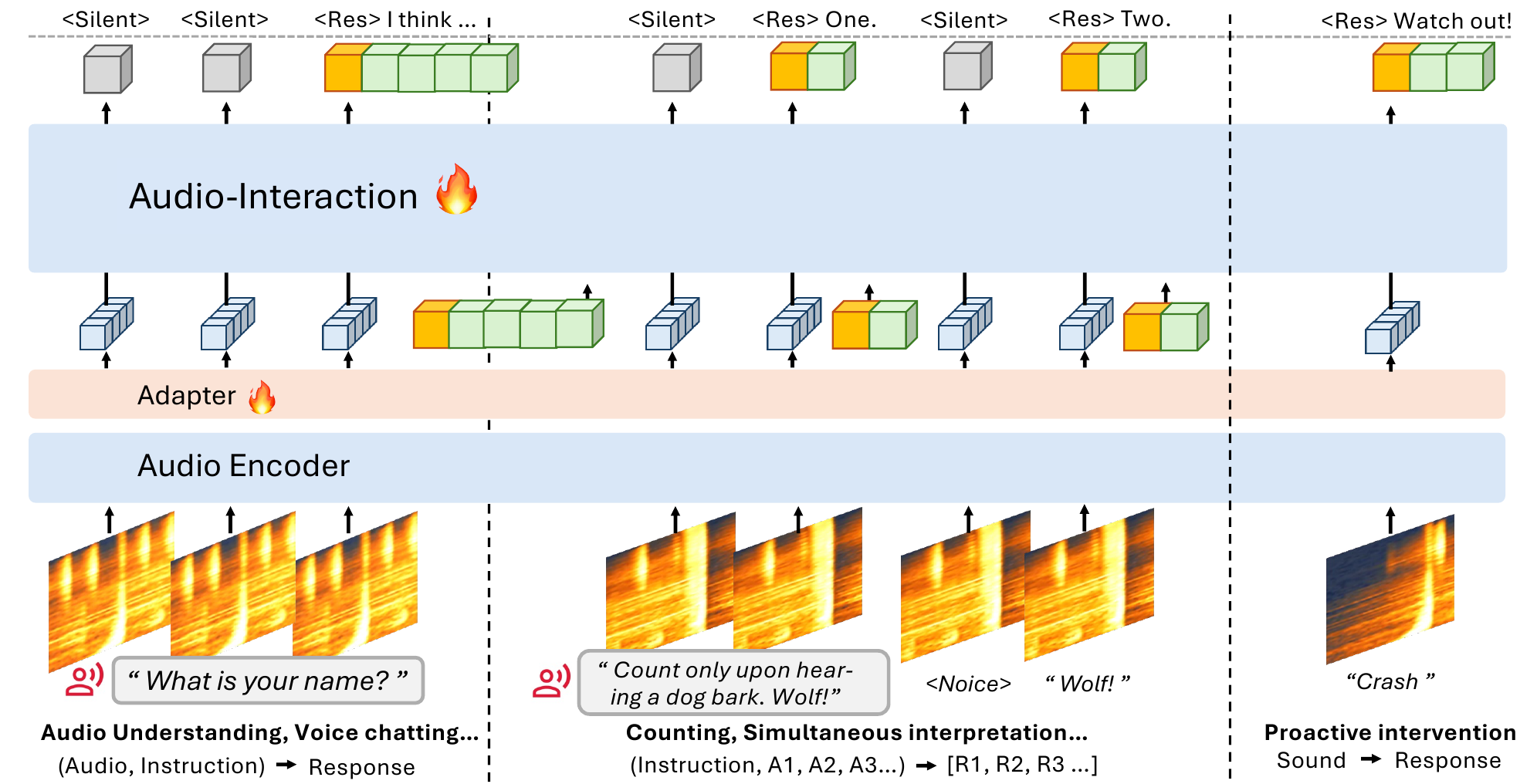}
\caption{\textsc{SoundFlow} training pipeline: streaming data and token-level supervision teach \textsc{Audio-Interaction} to perceive context, decide timing, and generate responses.}
\label{figure2-overview}
\end{figure}

\subsection{Streaming Training}

\textbf{Streaming sequence modeling.} Training and inference share the chunk-wise interface in Equation\,\ref{eq:pdr_loop} (Figure\,\ref{figure2-overview}). We use 400\,ms chunks to balance responsiveness and acoustic completeness. Each chunk updates $C_t$ before the model predicts $d_t\in\{\texttt{<silent>},\texttt{<response>}\}$; text is generated only for \texttt{<response>}. The model thus learns a single sequential process in which continuous perception supports grounded intervention timing and conditional response generation.

\textbf{Context memory and comprehension-aware silence training.} We supervise the two challenges jointly because C1 depends on C2. To address C2, \emph{history review} examples insert later questions about earlier turns, forcing the perceived state $C_t$ to retain long-range evidence despite noisy or semantically empty chunks. To address C1, comprehension-aware silence examples verified under the \textsc{Proactive-Sound-Bench} criteria teach the decision head to reject benign or irrelevant sounds. Together, they ensure that the model remembers the evidence and applies the correct intervention boundary to it.

\textbf{Dual-loss streaming conversion.} \textsc{Audio-Interaction} is initialized from Qwen2.5-Omni-3B, which provides a strong performance--efficiency trade-off for low-latency inference. The streaming objective directly trains the C1 silence/response decision, while the language modeling objective trains the response content. We optimize both jointly:
\[
\mathcal{L}
=
\frac{1}{N}\sum_{j=1}^{N}
\left(
\underbrace{-\log P_\theta\!\left(t_j \mid \mathcal{H}_j\right)}_{\mathcal{L}_{\mathrm{LM}}}
+
\lambda\underbrace{-\log P_\theta\!\left(s_j \mid \mathcal{H}_j\right)}_{\mathcal{L}_{\mathrm{stream}}}
\right),
\]
where $t_j$ denotes the target text token, $s_j$ denotes the target streaming control token, $\mathcal{H}_j$ denotes the decoding context, and $\lambda$ controls the relative weight of the decision objective. This coupling aligns \emph{decide} and \emph{respond} without sacrificing the general audio capabilities of the initialized model.

Let $\mathcal{A}^{\mathrm{ins}}$ denote the audio instruction, $\mathcal{A}^{\mathrm{in}}$ the input audio stream, and $\mathcal{T}$ the target response. Training proceeds in four stages. \emph{(1) Format training} teaches the target sequence format and the use of $\texttt{<Spe\_token>}$ on offline samples $(\mathcal{A}^{\mathrm{ins}}, \mathcal{A}^{\mathrm{in}} \rightarrow \mathcal{T})$.
\emph{(2) Adapter training} maps chunk-wise acoustic representations into the language-model space while keeping the format fixed.
\emph{(3) Large-scale streaming supervised training} jointly optimizes the adapter and language model on core capabilities, including audio understanding, ASR, speech translation, and spoken dialogue, using $(\mathcal{A}^{\mathrm{ins}} \rightarrow \mathcal{T})$ and $(\mathcal{A}^{\mathrm{ins}}, \mathcal{A}^{\mathrm{in}} \rightarrow \mathcal{T})$. 
\emph{(4) Instruction-following fine-tuning} introduces complex streaming behaviors such as continuous assistance, comprehension-aware intervention, and proactive response through interleaved multi-turn sequences such as $(\mathcal{A}^{\mathrm{ins}}, \mathcal{A}^{\mathrm{in}}_1, \mathcal{T}_1, \mathcal{A}^{\mathrm{in}}_2, \mathcal{T}_2, \ldots)$, $(\mathcal{A}^{\mathrm{ins}}, \mathcal{A}^{\mathrm{in}}_1, \mathcal{A}^{\mathrm{in}}_2, \mathcal{T}, \mathcal{A}^{\mathrm{in}}_3, \mathcal{T}, \ldots)$, and $(\mathcal{A}^{\mathrm{in}} \rightarrow \mathcal{T})$.

\begin{figure}[!t]
\centering
\includegraphics[width=0.9\linewidth]{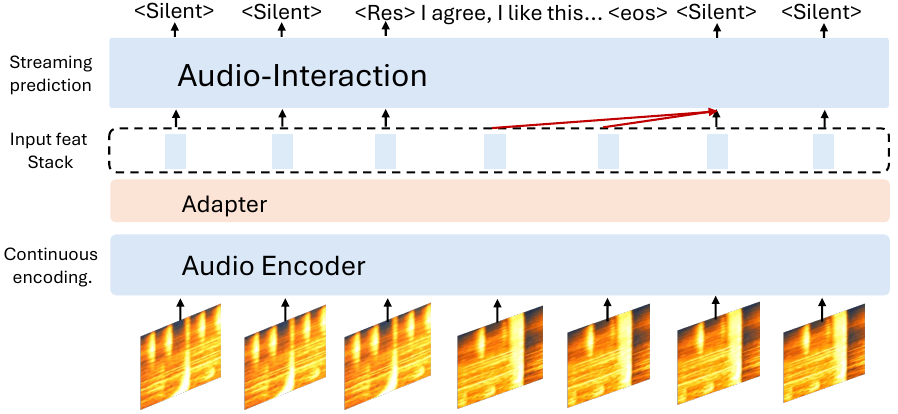}
\caption{\textsc{SoundFlow} asynchronous FIFO inference queues audio chunks in time order while decoding runs only when the model is not speaking.}
\label{fig:fifo_inference}
\end{figure}

\subsection{Stabilizing Asynchronous Inference via FIFO Scheduling.}
C2 also arises at deployment: autoregressive response generation can block audio encoding and break the context needed for the next C1 decision. To prevent this, \textsc{SoundFlow} uses asynchronous FIFO scheduling. In Figure\,\ref{fig:fifo_inference}, the encoder processes incoming chunks and appends their projected acoustic representations to a temporally ordered queue. At step $t$, the incoming chunk $x_t$ is encoded into $\mathbf{a}_t$ and appended to $\mathcal{Q}_t$. The decoder is triggered only when the last generated token $r_{t-1}\in\{\texttt{<eos>},\texttt{<silent>}\}$; it then consumes queued features in order and produces the next control token or response. Otherwise, decoding continues autoregressively while the encoder keeps listening. This decoupling preserves the \emph{perceive} state during \emph{respond}, maintains temporal order, and enables non-blocking listening, which reduces average first-chunk latency and prevents streaming stalls in Table~\ref{tab:fifo_inference}.

\begin{table*}[t]
\centering
\setlength{\tabcolsep}{3pt}
\renewcommand{\arraystretch}{1.0}

\begin{tabular}{@{}l c c c cccc cccc@{}}
\toprule
\multirow{2.5}{*}{\textbf{Model}}
  & \multirow{2.5}{*}{\textbf{Size}}
  & \multirow{2.5}{*}{\textbf{Stream.}}
  & \multirow{2.5}{*}{\textbf{Multi-turn}}
  & \multicolumn{4}{c}{\textbf{Text instruction}}
  & \multicolumn{4}{c}{\textbf{Audio instruction}} \\
\cmidrule(lr){5-8}
\cmidrule(lr){9-12}
  & & & & Sound & Music & Speech & Avg.
  & Sound & Music & Speech & Avg. \\
\midrule

\multicolumn{12}{@{}l}{\textit{\textbf{Large Audio Language Models}}} \\
Audio Flamingo 2
  & 3B
  & \ding{55}
  & \ding{55}
  & \textbf{71.47}
  & \textbf{70.96}
  & 44.74
  & \underline{62.40}
  & 1.50
  & 1.49
  & 0.35
  & 1.16 \\

Qwen2-Audio
  & 7B
  & \ding{55}
  & \ding{51}
  & 54.95
  & 50.98
  & 42.04
  & 49.20
  & 22.32
  & 19.16
  & 16.31
  & 19.41 \\

Voxtral-Mini
  & 3B
  & \ding{55}
  & \ding{51}
  & 58.56
  & 49.70
  & 43.53
  & 50.60
  & 46.08
  & 34.13
  & 30.50
  & 37.24 \\

Audio-Reasoner
  & 8.4B
  & \ding{55}
  & \ding{55}
  & 60.06
  & 64.30
  & \textbf{60.70}
  & 61.71
  & 20.48
  & 26.65
  & 13.48
  & 20.57 \\

\multicolumn{12}{@{}l}{\textit{\textbf{Omni Language Models}}} \\
Qwen2.5-Omni
  & 3B
  & \ding{55}
  & \ding{51}
  & 65.36
  & 48.94
  & 57.78
  & 57.81
  & 51.81
  & 44.01
  & 29.79
  & 42.51 \\

Qwen2.5-Omni
  & 7B
  & \ding{55}
  & \ding{51}
  & \underline{67.87}
  & \underline{69.16}
  & \underline{59.76}
  & \textbf{65.60}
  & \underline{60.54}
  & \underline{50.90}
  & \underline{35.11}
  & \underline{49.58} \\

Phi-4-multimodal
  & 5.6B
  & \ding{55}
  & \ding{51}
  & 60.97
  & 52.87
  & 52.83
  & 55.56
  & 44.65
  & 27.84
  & 21.99
  & 31.75 \\

Baichuan-Omni-1.5
  & 7B
  & \ding{55}
  & \ding{51}
  & 65.47
  & 58.98
  & 55.26
  & 59.90
  & 57.53
  & 36.53
  & 24.82
  & 40.40 \\

\multicolumn{12}{@{}l}{\textit{\textbf{Streaming Audio Language Models}}} \\
\textbf{Audio-Interaction}
  & 3B
  & \ding{51}
  & \ding{51}
  & 64.12
  & 47.80
  & 55.13
  & 55.68
  & \best{65.63}
  & \best{57.93}
  & \best{46.68}
  & \best{58.15} \\
\bottomrule
\end{tabular}

\caption{MMAU accuracy under text and spoken instructions across sound, music, and speech. \textbf{Stream.} and \textbf{Multi-turn} indicate streaming and multi-turn training support (\texttt{-}: not applicable).}
\label{tab:mmau}
\end{table*}
\section{StreamAudio-2M Dataset}
\label{sec:data_section}

\subsection{Overview}
\label{sec:data_overview}

\textit{StreamAudio-2M} supports \textsc{SoundFlow} with long-form supervision absent from conventional \textit{(clip, instruction, response)} datasets~\citep{kong2024audioflamingo,chu2024qwen2}: continuous acoustic context, sparse \texttt{<silent>}/\texttt{<response>} labels, and task-specific outputs. It contains \textbf{2.6M} items (\textbf{302k hours}), each comprising a \textbf{3--15 turn} heterogeneous interaction with interleaved events and context-dependent response cues. In supplementary Figures\,\ref{fig:streamaudio_main} and\,\ref{fig:overview} and Table\,\ref{tab:overview_sources}, the corpus spans \textbf{7 categories}---\textit{Audio Agent, Proactive Respond, Voice Chatting, Streaming Audio Understanding, Following Music, Real-time ASR}, and \textit{Streaming Translation}---and \textbf{28 sub-tasks}.

\subsection{Curation Pipeline}
\label{sec:curation_pipeline}

The pipeline includes: \textbf{(i) Data Collection.} We gather $\sim$1.64M foundational items ($\sim$8{,}900 hours) from dialogue (MOSS), ASR (CommonVoice, GigaSpeech, LibriSpeech~\citep{panayotov2015librispeech}, VoxPopuli), speech translation (CoVoST2~\citep{wang2021covost}, AISHELL), and music/audio-QA sources (FMA, AudioSet~\citep{gemmeke2017audio}). We augment them with $\sim$171k acoustic-event clips (AudioSet, AudioX~\citep{tian2025audiox}, ElevenLabs) and environmental noise from MUSAN~\citep{snyder2015musan}, WHAM!~\citep{wichern2019wham}, and DNS-Challenge~\citep{timcheck2023intel}. \textbf{(ii) Preprocessing.} Text is synthesized with multi-voice CosyVoice and verified through LLM rewriting and ASR checking, while TFJP normalizes audio boundaries. \textbf{(iii) Sequence Concatenation.} The hierarchical event-selection strategy in Section\,\ref{sec3.2: Streaming Data Construction} composes validated instances into coherent streams with dual-track environmental noise. \textbf{(iv) Token-level Annotation.} Each stream is converted into $\langle\text{input ids}, \text{labels}\rangle$ pairs containing aligned \texttt{<silent>}/\texttt{<response>} labels and target text for streaming training.

\subsection{Proactive-Sound-Bench}
\label{sec:proactive_bench}
 
\textbf{Proactive-Sound-Bench} evaluates proactive streaming response through \textbf{644} human-designed acoustic events, each requiring the model to trigger or abstain within a continuous stream. The benchmark spans six top-level categories and 17 subcategories. The \emph{Single} tier tests isolated decisions, while the \emph{Multiple} tier concatenates same-category events to test sustained intervention under distractors. We report average accuracy; Table\,\ref{tab_meso_definitions} provides per-category definitions.

\begin{table*}[t]
\centering

\begin{minipage}[t]{0.49\textwidth}
\centering
\setlength{\tabcolsep}{3.5pt}
\renewcommand{\arraystretch}{1.0}

\begin{tabular*}{\linewidth}{@{\extracolsep{\fill}}l c c c c c@{}}
\toprule
\multirow{2.5}{*}{\textbf{Model}}
  & \multirow{2.5}{*}{\textbf{Size}}
  & \multicolumn{2}{c}{\textbf{SpokenQA}}
  & \multicolumn{2}{c}{\textbf{Voicebench}} \\
\cmidrule(lr){3-4}
\cmidrule(lr){5-6}
  & & LLa. Q. & Web Q. & Alpa. & SD-QA \\
\midrule

\multicolumn{6}{@{}l}{\textit{\textbf{Specialized Models}}} \\
Moshi
  & 7B
  & 62.20
  & 26.30
  & 2.01
  & 15.01 \\

Freeze-Omni
  & 7B
  & 72.00
  & 44.73
  & 4.14
  & 50.16 \\

\multicolumn{6}{@{}l}{\textit{\textbf{Omni \& Audio Language Models}}} \\
Baichuan-Omni-1.5
  & 7B
  & \textbf{78.50}
  & \underline{59.10}
  & \textbf{4.50}
  & 43.40 \\

Qwen2-Audio
  & 7B
  & 69.67
  & 45.20
  & 3.74
  & 35.71 \\

Qwen2.5-Omni
  & 3B
  & 66.00
  & 27.95
  & 4.32
  & 49.37 \\

Qwen2.5-Omni
  & 7B
  & \underline{75.33}
  & \textbf{62.80}
  & \underline{4.49}
  & \textbf{55.71} \\

Phi-4-multimodal
  & 5.6B
  & 60.2
  & 26.6
  & 3.81
  & 39.78 \\

\multicolumn{6}{@{}l}{\textit{\textbf{Streaming Audio Language Models}}} \\
\textbf{Audio-Interaction}
  & 3B
  & 67.31
  & 54.34
  & 4.28
  & \underline{52.14} \\
\bottomrule
\end{tabular*}

\caption{Performance score ($\uparrow$) on four spoken-dialogue benchmarks.}
\label{tab:dialogue}
\end{minipage}
\hfill
\begin{minipage}[t]{0.49\textwidth}
\centering
\setlength{\tabcolsep}{3pt}
\renewcommand{\arraystretch}{1.0}

\begin{tabular*}{\linewidth}{@{\extracolsep{\fill}}l c c c c c@{}}
\toprule
\multirow{2.5}{*}{\textbf{Model}}
  & \multirow{2.5}{*}{\textbf{Size}}
  & \multicolumn{2}{c}{\textbf{LibriSpeech}}
  & \multicolumn{2}{c}{\textbf{CoVoST2}} \\
\cmidrule(lr){3-4}
\cmidrule(lr){5-6}
  & & clean & other & en-zh & zh-en \\
\midrule

\multicolumn{6}{@{}l}{\textit{\textbf{Specialized Models}}} \\
Canary
  & 1B
  & \textbf{1.48}
  & \textbf{2.93}
  & -
  & - \\

Canary-Qwen
  & 2.5B
  & \underline{1.49}
  & \underline{3.10}
  & -
  & - \\

\multicolumn{6}{@{}l}{\textit{\textbf{Omni \& Audio Language Models}}} \\
Baichuan-Omni-1.5
  & 7B
  & 5.71
  & 10.09
  & -
  & - \\

Qwen2-Audio
  & 7B
  & 1.60
  & 3.60
  & 45.20
  & 24.40 \\

Qwen2.5-Omni
  & 3B
  & 2.87
  & 5.90
  & 39.50
  & 18.17 \\

Qwen2.5-Omni
  & 7B
  & 1.80
  & 3.40
  & 41.40
  & \underline{29.40} \\

Phi-4-multimodal
  & 5.6B
  & 1.69
  & 3.82
  & \underline{46.30}
  & 22.39 \\

\multicolumn{6}{@{}l}{\textit{\textbf{Streaming Audio Language Models}}} \\
\textbf{Audio-Interaction}
  & 3B
  & 3.17
  & 6.04
  & \best{55.22}
  & \best{35.21} \\
\bottomrule
\end{tabular*}

\caption{WER (\%, $\downarrow$) on LibriSpeech and speech-to-text translation BLEU ($\uparrow$) on CoVoST2.}
\label{tab:asr}
\end{minipage}

\end{table*}

\begin{table*}[t]
\centering
\setlength{\tabcolsep}{4.2pt}
\renewcommand{\arraystretch}{1.1}

\begin{tabular}{@{}l!{\vrule width 0.4pt}cccccccccccc!{\vrule width 0.4pt}cc@{}}
\toprule
\multirow{2.5}{*}{\textbf{Model}}
  & \multicolumn{2}{c}{\textbf{Human}}
  & \multicolumn{2}{c}{\textbf{Daily}}
  & \multicolumn{2}{c}{\textbf{Equip.}}
  & \multicolumn{2}{c}{\textbf{Traffic}}
  & \multicolumn{2}{c}{\textbf{Nature}}
  & \multicolumn{2}{c!{\vrule width 0.4pt}}{\textbf{Music}}
  & \multicolumn{2}{c}{\textbf{Avg.}} \\
\cmidrule(lr){2-3}
\cmidrule(lr){4-5}
\cmidrule(lr){6-7}
\cmidrule(lr){8-9}
\cmidrule(lr){10-11}
\cmidrule(lr){12-13}
\cmidrule(lr){14-15}
  & Sin. & Mul.
  & Sin. & Mul.
  & Sin. & Mul.
  & Sin. & Mul.
  & Sin. & Mul.
  & Sin. & Mul.
  & Sin. & Mul. \\
\midrule

\multicolumn{15}{@{}l}{\textit{\textbf{Omni \& Audio Language Models}}} \\
Qwen2.5-Omni-3B
  & 37.2 & 28.9
  & 48.1 & 42.5
  & 30.0 & 17.9
  & 44.9 & 36.7
  & 45.6 & 17.5
  & 53.3 & 40.0
  & 41.0 & 29.3 \\

Qwen2.5-Omni-7B
  & \underline{54.5} & 34.6
  & \underline{72.9} & 40.2
  & 47.9 & 19.3
  & 53.1 & 24.5
  & \underline{55.3} & 31.1
  & 53.3 & \textbf{60.0}
  & 58.2 & 32.1 \\

Kimi-Audio-Instruct
  & 39.1 & 26.3
  & 61.3 & 38.6
  & 28.6 & 22.1
  & 28.6 & 16.3
  & 26.2 & 28.2
  & 26.7 & 26.7
  & 39.9 & 28.4 \\

MiniCPM-o-4.5
  & 53.8 & 53.2
  & \textbf{75.1} & \textbf{75.4}
  & \underline{52.9} & \underline{52.9}
  & \underline{55.1} & \underline{55.1}
  & 48.5 & 47.6
  & 53.3 & 53.3
  & \underline{58.9} & \underline{58.9} \\

Step-Audio 2
  & 9.6 & 5.8
  & 7.7 & 3.4
  & 4.3 & 0.0
  & 12.2 & 6.1
  & 14.6 & 1.0
  & 6.7 & 0.0
  & 8.9 & 3.0 \\

Gemini-3-Flash
  & 48.1 & \underline{59.6}
  & 32.0 & 47.5
  & 25.7 & 40.0
  & 28.6 & 53.1
  & 48.5 & \underline{56.3}
  & 33.3 & 53.3
  & 37.0 & 50.8 \\

\multicolumn{15}{@{}l}{\textit{\textbf{Streaming Audio Language Models}}} \\
\textbf{Audio-Interaction}
  & \textbf{56.4} & \textbf{64.9}
  & 68.1 & \underline{65.8}
  & \textbf{57.1} & \textbf{55.7}
  & \textbf{64.9} & \textbf{69.0}
  & \textbf{61.8} & \textbf{61.8}
  & \textbf{66.7} & \textbf{60.0}
  & \textbf{61.2} & \textbf{62.8} \\
\bottomrule
\end{tabular}

\caption{Results on the Proactive-Sound-Bench. \textit{Equip.} stands for Equipment. \textbf{Sin.} and \textbf{Mul.} denote Single-round and Multi-round, respectively. \textbf{Best} and \underline{second-best} results are highlighted.}
\label{tab:proactive-sound-bench}
\end{table*}

\section{Experiments}
\label{sec:experiments}

\subsection{Settings}
\label{sec:settings}

Our experiments address three questions derived from the \emph{perceive--decide--respond} formulation. \textbf{(RQ1) Capability preservation:} does streaming conversion retain the general audio and language abilities of an offline LALM? We evaluate MMAU~\citep{sakshi2024mmau}, four spoken-dialogue benchmarks---AlpacaEval~\citep{dubois2023alpacafarm}, SD-QA~\citep{faisal2021sd}, Llama Questions~\citep{nachmani2023spoken}, and Web Questions~\citep{berant2013semantic} under VoiceBench~\citep{chen2026voicebench}---LibriSpeech~\citep{panayotov2015librispeech}, and CoVoST2~\citep{wang2021covost}. \textbf{(RQ2/C1) Grounded decisions:} can the model determine when to respond or abstain from accumulated acoustic meaning? Proactive-Sound-Bench directly evaluates this capability. \textbf{(RQ3/C2) Continuous interaction:} can the model preserve context across chunks and continue listening during generation? We assess long-stream stability, internal cross-chunk continuity, and asynchronous inference latency.

Baselines cover \textbf{Audio LLMs} (Audio Flamingo 2~\citep{ghosh2025audio}, Qwen2-Audio~\citep{chu2024qwen2}, Voxtral-Mini~\citep{liu2025voxtral}, and Audio-Reasoner~\citep{audio-reasoner}), \textbf{Omni LLMs} (Qwen2.5-Omni~\citep{qwen25omni2025}, Baichuan-Omni-1.5~\citep{li2025baichuan}, and Phi-4-multimodal~\citep{abouelenin2025phi}), and \textbf{task-specialized models} (Whisper-large-v3~\citep{radford2023whisper} and Canary~\citep{sekoyan2025canary} for ASR; Moshi~\citep{defossez2024moshi}, Freeze-Omni~\citep{wang2024freeze}, and LLaMA-Omni2~\citep{fang2025llama} for streaming dialogue).

\subsection{Main Results}
\label{sec:main_results}


\textit{\textbf{RQ1: Streaming conversion preserves broad audio capability.}} On MMAU (Table\,\ref{tab:mmau}), \textsc{Audio-Interaction} remains close to its Qwen2.5-Omni-3B initialization under text instructions (55.68 vs. 57.81) while improving the audio-instruction average from 42.51 to \textbf{58.15}. This indicates that streaming training preserves general understanding and substantially improves robustness to spoken instructions. On dialogue (Table\,\ref{tab:dialogue}), it improves over the initialization by 1.31 points on Llama Questions, 26.39 on Web Questions, and 2.77 on SD-QA, with only a 0.04 decrease on AlpacaEval. On CoVoST2 (Table\,\ref{tab:asr}), it gains \textbf{15.72/17.04} BLEU on En--Zh/Zh--En and approaches or exceeds larger baselines. LibriSpeech WER increases modestly from 2.87/5.90 to 3.17/6.04 after replacing utterance-level decoding with chunk-wise streaming, showing a limited cost for continuous operation.

\begin{figure}[!t]
\centering
\includegraphics[width=\linewidth]{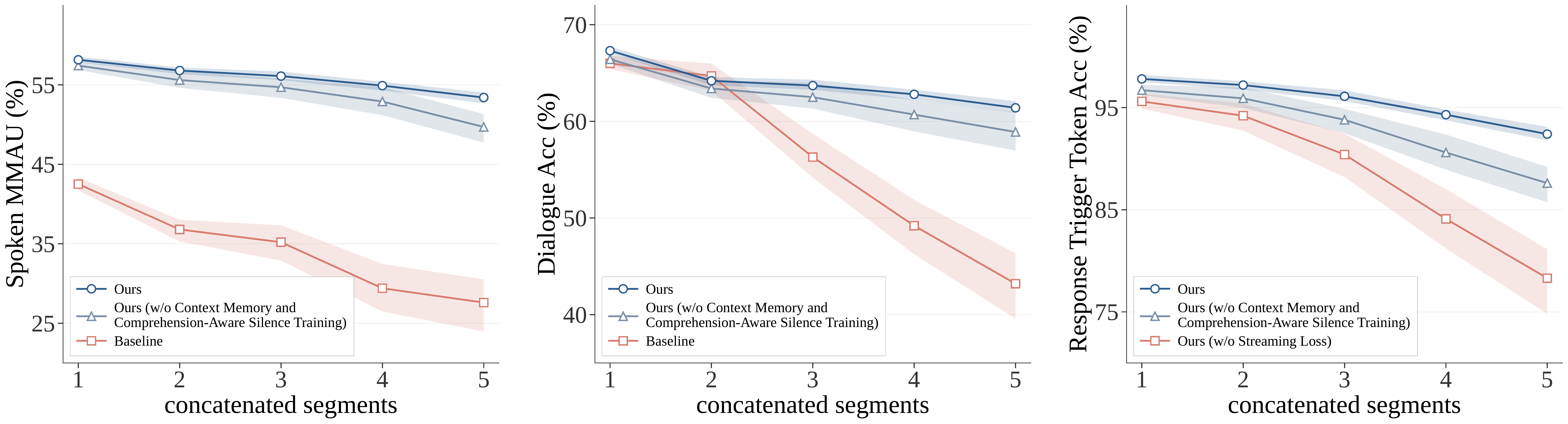}
\caption{Stability of \textsc{Audio-Interaction} as the stream extends from 1 to 5 concatenated segments. We report MMAU average accuracy, dialogue accuracy, and end-to-end latency.}
\label{fig:ablation_stability}
\end{figure}

\textit{\textbf{RQ2/C1: Decisions are grounded in evolving acoustic context.}} Proactive-Sound-Bench requires both intervention on assistance-worthy events and abstention on benign distractors. \textsc{Audio-Interaction} achieves the best average results in both the Single (\textbf{61.2}) and Multi (\textbf{62.8}) tiers (Table\,\ref{tab:proactive-sound-bench}). Its stronger Multi result shows the \emph{decide} stage remains reliable as events accumulate, rather than reacting only to isolated local cues.

\textit{\textbf{RQ3/C2: Continuous perception remains stable over long streams.}} As the number of concatenated segments grows from one to five, \textsc{Audio-Interaction} retains more than 91\% of its single-segment accuracy, whereas the offline baseline loses over 30\% (Figure\,\ref{fig:ablation_stability}). This result shows that the learned context supports repeated perceive--decide--respond cycles instead of degrading after each chunk or interaction.

\begin{figure}[t]
\centering
\includegraphics[width=\linewidth]{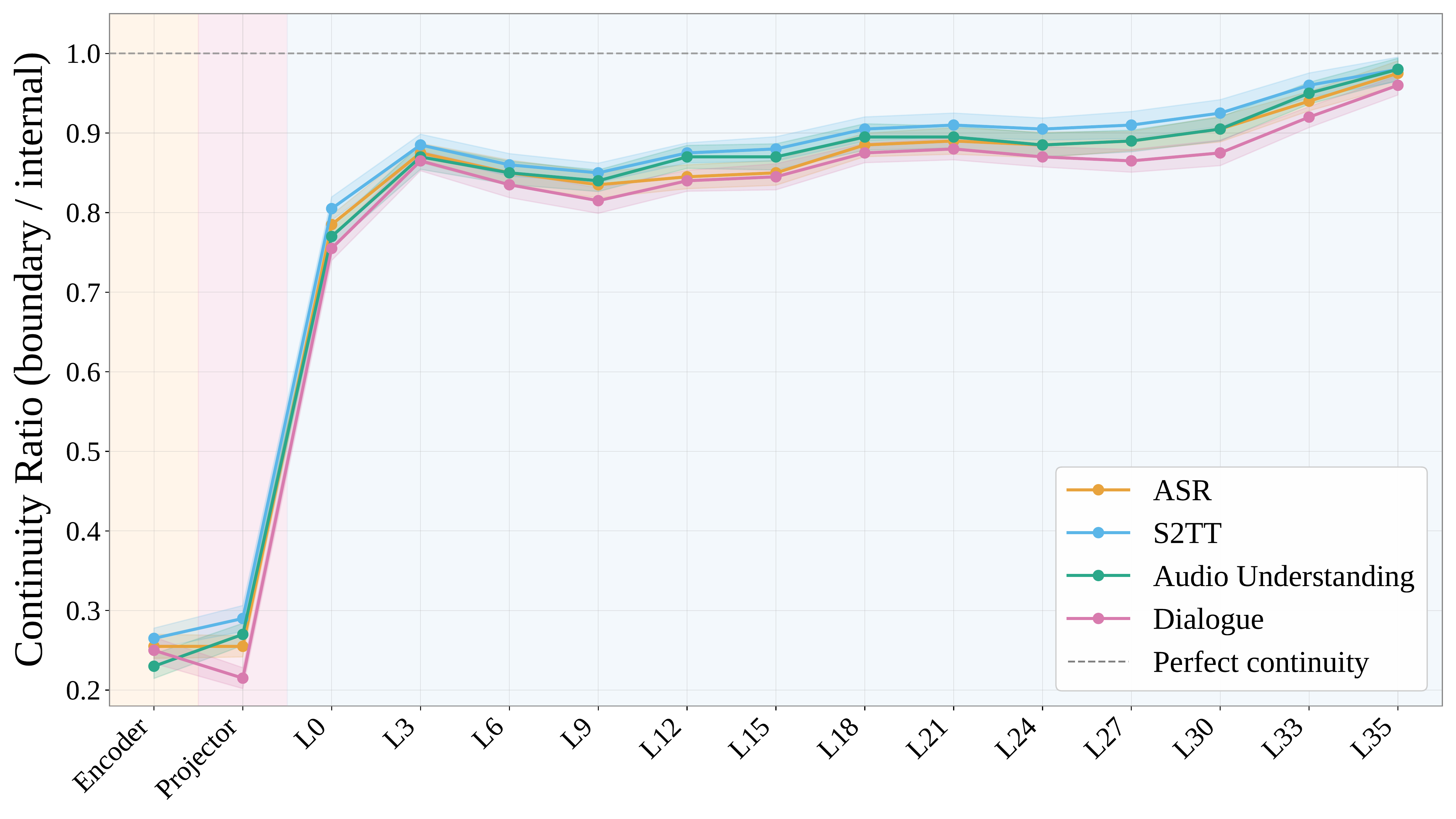}
\caption{Cross-chunk continuity ratio across the audio encoder, projector, and GPT blocks on four tasks; higher values indicate smoother boundary representations.}
\label{fig:continuity}
\end{figure}

\begin{figure}[t]
\centering
\includegraphics[width=\linewidth]{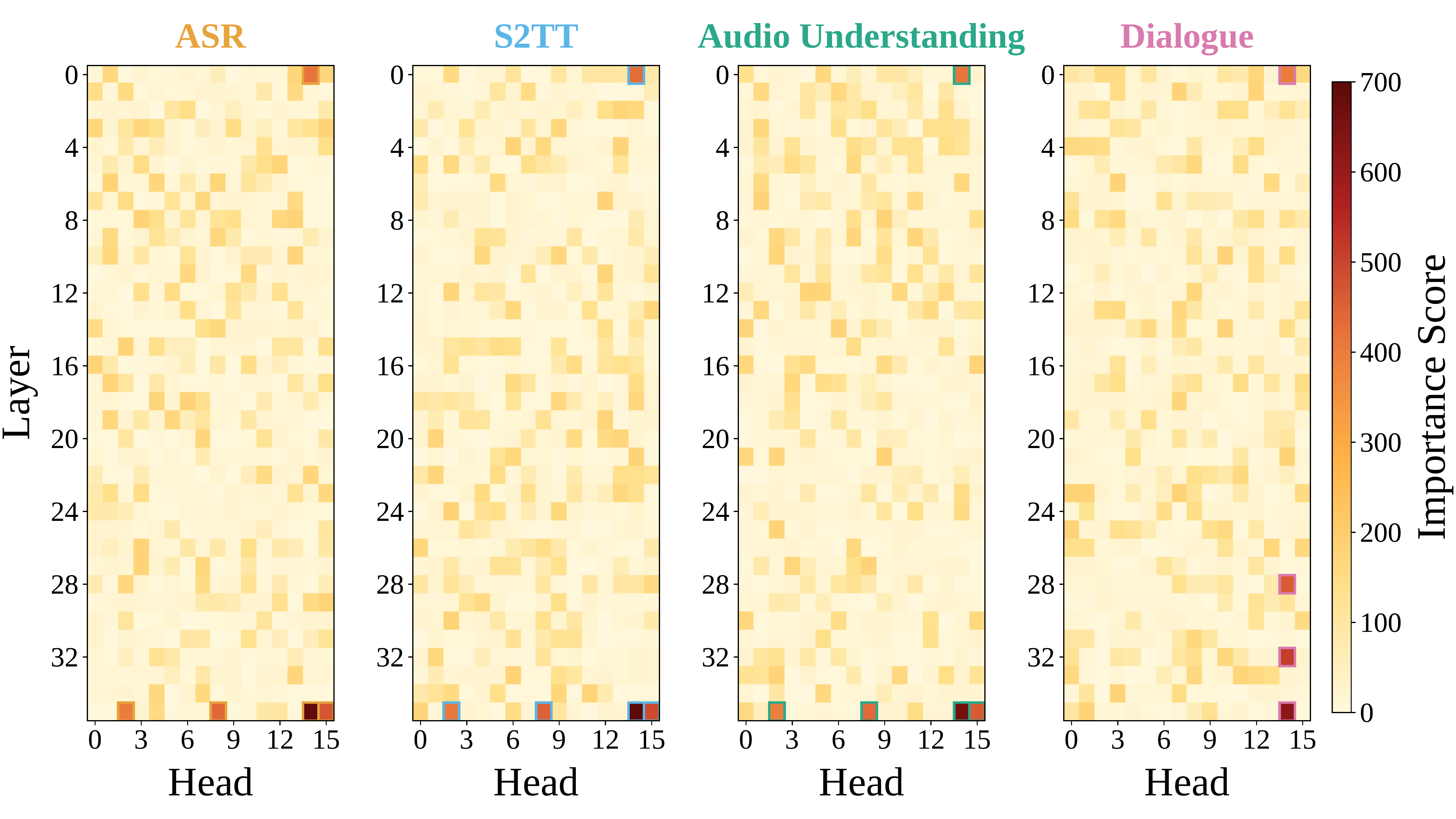}
\caption{Attention-head importance for streaming control-token generation, measured by performance degradation under single-head ablation.}
\label{fig:head_importance}
\end{figure}

\begin{table*}[t]
\centering
\begin{minipage}[t]{0.48\textwidth}
\centering
\setlength{\tabcolsep}{3pt}
{\small
\begin{tabularx}{\linewidth}{@{}l >{\raggedright\arraybackslash}X c c c@{}}
\toprule
\textbf{Variant}
& \textbf{Configuration}
& \textbf{MMAU}$\uparrow$
& \textbf{Alpaca.}$\uparrow$
& \textbf{Trig. Acc.}$\uparrow$ \\
\midrule
V1 & Baseline & 42.51 & \textbf{4.32} & -- \\
\midrule
V2 & + Streaming SFT & \textbf{58.56} & 4.17 & 92.42\% \\
V3 & V2 w/o TFJP pre. & 57.74 & 4.19 & 85.35\% \\
V4 & V2 w/o Event sel. & 55.11 & 4.25 & 88.51\% \\
\midrule
V5 & \textbf{Audio-Interaction} & 58.15 & 4.28 & \textbf{96.77}\% \\
\bottomrule
\end{tabularx}
}
\addtocounter{table}{1}
\caption{Streaming data and training components studies.}
\label{tab:ablation_data}
\end{minipage}
\hfill
\begin{minipage}[t]{0.48\textwidth}
\centering
\setlength{\tabcolsep}{6pt}
{\small
\begin{tabularx}{\linewidth}{@{}>{\raggedright\arraybackslash}X c c c@{}}
\toprule
\textbf{Variant}
& \textbf{Alpaca.}$\uparrow$
& \textbf{MMAU}$\uparrow$
& \textbf{Lat.}$\downarrow$ \\
\midrule
Baseline & 4.32 & 57.81 & -- \\
\midrule
Chunk = 0.2\,s & 3.41 & 49.74 & \textbf{258} \\
Chunk = 0.6\,s & 4.27 & \underline{58.46} & 674 \\
Chunk = 0.8\,s & \textbf{4.30} & \textbf{59.13} & 786 \\
\midrule
\textbf{Chunk = 0.4\,s} & \underline{4.28} & 58.15 & \underline{392} \\
\bottomrule
\end{tabularx}
}
\caption{Effect of chunk size.}
\label{tab:ablation_chunk}
\end{minipage}
\end{table*}

\subsection{Additional Analysis}

We test if the model internally implements the two mechanisms required by the formulation: continuous context for C2 and a shared silence/response policy for C1.

\textit{\textbf{[C2/Perceive] The early decoder reconstructs cross-chunk continuity.}} Each 0.4\,s chunk is encoded independently, with separate positional embeddings and no cross-chunk encoder attention. We measure the resulting fragmentation using a \emph{continuity ratio}: boundary-pair cosine similarity normalized by intra-chunk similarity, where 1.0 denotes seamless continuity. In Figure\,\ref{fig:continuity}, the ratio is 0.25 at the encoder and changes by $<$0.02 after projection, but rises to 0.80 at GPT Layer~0. The shared trend across four tasks shows cross-chunk KV-cache access reconstructs the continuous context $C_t$ at the start of the decoder, providing the representational basis for C2.

\begin{table}[t]
\centering
\setlength{\tabcolsep}{6pt}
\renewcommand{\arraystretch}{1.0}
\begin{tabular*}{\columnwidth}{@{\extracolsep{\fill}}lcc@{}}
\toprule
\textbf{Settings} & \textbf{Avg. FCL} & \textbf{Stall \%} \\
\midrule
\textsc{Ours} & \textbf{392ms} & \textbf{0.0\%} \\
\quad $w/o$ FIFO & 831ms & 5.2\% \\
\bottomrule
\end{tabular*}
\addtocounter{table}{-3}
\caption{Effect of asynchronous FIFO inference on first-chunk latency (FCL) and streaming stalls.}
\label{tab:fifo_inference}
\end{table}
\addtocounter{table}{2}

\textit{\textbf{[C1/Decide] A shared attention head mediates silence/response control.}} We ablate each attention head and measure the degradation in \texttt{<silent>}/\texttt{<response>} prediction. L35H14 dominates across all four tasks; ablating it alone reduces the S2TT token-match score by 0.88 (Figure\,\ref{fig:head_importance}). The same head governs different tasks, suggesting that C1 is implemented as a shared interaction policy over accumulated context rather than as separate task-specific triggers.

\subsection{Ablation Study}

Ablations examine how data construction, chunk size, dual-loss training, and FIFO inference support the \emph{perceive--decide--respond} loop and address C1/C2.

\textbf{[Respond/C2] FIFO inference prevents generation from interrupting perception.} Removing FIFO scheduling increases average first-chunk latency from $392$\,ms to $831$\,ms ($2.12\times$) and the stall rate from $0.0\%$ to $5.2\%$ (Table\,\ref{tab:fifo_inference}). Thus, decoupling encoding from decoding is necessary to keep updating context while a response is generated.

\textbf{[Perceive--Decide/C2--C1] Streaming data quality determines grounded triggering.} Streaming SFT (V2) raises MMAU from $42.51$ to $58.56$ and reaches $92.42\%$ trigger accuracy (Table\,\ref{tab:ablation_data}). Removing TFJP (V3) lowers trigger accuracy by 7.07 points, showing that boundary artifacts disrupt reliable decisions. Removing hierarchical event selection (V4) lowers MMAU by 3.45 points and trigger accuracy by 3.91 points, confirming that semantically coherent histories support both context construction and decision grounding. The full model (V5) achieves the best balance, recovering AlpacaEval to 4.28 and raising trigger accuracy to $96.77\%$.

\textbf{[Perceive/C2] Chunk size balances context completeness and responsiveness.} A 0.2\,s chunk reduces AlpacaEval/MMAU to 3.41/49.74 because each update contains insufficient semantic evidence (Table\,\ref{tab:ablation_chunk}). Chunks of 0.6--0.8\,s recover accuracy but increase latency to 674--786\,ms. The 0.4\,s setting retains competitive accuracy (4.28/58.15) at 392\,ms, balancing coherent perception with timely decisions.

\textbf{[Decide--Respond] Dual-loss weighting balances control and content.} Increasing $\lambda$ improves trigger accuracy from $95.3\%$ to $96.9\%$, but $\lambda{=}2.0$ reduces MMAU to 57.3 (Table\,\ref{tab:ablation_lambda}). We therefore use $\lambda{=}1.0$, which retains 58.2 MMAU while reaching $96.7\%$ trigger accuracy, balancing C1 decision learning with response capability.

\begin{table}[t]
\centering
\small
\setlength{\tabcolsep}{6pt}
\renewcommand{\arraystretch}{1.0}
\begin{tabular*}{\columnwidth}{@{\extracolsep{\fill}}lccc@{}}
\toprule
$\lambda$               & $0.5$  & $1.0$  & $2.0$  \\
\midrule
MMAU$\uparrow$          &\textbf{ 58.3 }  & 58.2   & 57.3   \\
Trigger Acc.$\uparrow$   & 95.3   & \textbf{96.7}   & 96.9   \\
\bottomrule
\end{tabular*}
\caption{Effect of dual-loss weight $\lambda$.}
\label{tab:ablation_lambda}
\end{table}

\section{Conclusion}

We formalized the \textbf{\textsc{Audio Interaction Model}} as an always-on \emph{perceive--decide--respond} paradigm and instantiated it with \textbf{\textsc{Audio-Interaction}}. The formulation connects two coupled requirements: preserving coherent context across chunks and concurrent generation (C2), and grounding silence/response decisions in that context (C1). \textbf{\textsc{SoundFlow}} addresses them through streaming-native data construction and history-aware training for continuous perception, comprehension-aware supervision for response decisions, and dual-loss training with asynchronous FIFO inference for non-blocking generation. We further introduce \textbf{\textsc{StreamAudio-2M}}, a 2.6M-item, 302k-hour corpus, and \textbf{\textsc{Proactive-Sound-Bench}}. Across 8 benchmarks, \textsc{Audio-Interaction} remains competitive on conventional audio tasks while enabling spoken-instruction robustness, stable long-stream interaction, and proactive intervention. These results establish audio interaction as a unified extension of offline LALMs toward continuously listening, context-aware audio intelligence.



\bibliography{aaai2027}

@article{chu2023qwen,
  title={Qwen-audio: Advancing universal audio understanding via unified large-scale audio-language models},
  author={Chu, Yunfei and Xu, Jin and Zhou, Xiaohuan and Yang, Qian and Zhang, Shiliang and Yan, Zhijie and Zhou, Chang and Zhou, Jingren},
  journal={arXiv preprint arXiv:2311.07919},
  year={2023}
}

@article{tang2023salmonn,
  title={Salmonn: Towards generic hearing abilities for large language models},
  author={Tang, Changli and Yu, Wenyi and Sun, Guangzhi and Chen, Xianzhao and Tan, Tian and Li, Wei and Lu, Lu and Ma, Zejun and Zhang, Chao},
  journal={arXiv preprint arXiv:2310.13289},
  year={2023}
}

@article{chu2024qwen2,
  title={Qwen2-audio technical report},
  author={Chu, Yunfei and Xu, Jin and Yang, Qian and Wei, Haojie and Wei, Xipin and Guo, Zhifang and Leng, Yichong and Lv, Yuanjun and He, Jinzheng and Lin, Junyang and others},
  journal={arXiv preprint arXiv:2407.10759},
  year={2024}
}

@article{xie2024mini,
  title={Mini-omni: Language models can hear, talk while thinking in streaming},
  author={Xie, Zhifei and Wu, Changqiao},
  journal={arXiv preprint arXiv:2408.16725},
  year={2024}
}

@article{defossez2024moshi,
  title={Moshi: a speech-text foundation model for real-time dialogue},
  author={D{\'e}fossez, Alexandre and Mazar{\'e}, Laurent and Orsini, Manu and Royer, Am{\'e}lie and P{\'e}rez, Patrick and J{\'e}gou, Herv{\'e} and Grave, Edouard and Zeghidour, Neil},
  journal={arXiv preprint arXiv:2410.00037},
  year={2024}
}

@article{fang2024llama,
  title={Llama-omni: Seamless speech interaction with large language models},
  author={Fang, Qingkai and Guo, Shoutao and Zhou, Yan and Ma, Zhengrui and Zhang, Shaolei and Feng, Yang},
  journal={arXiv preprint arXiv:2409.06666},
  year={2024}
}

@article{qwen25omni2025,
  title   = {{Qwen2.5-Omni} Technical Report},
  author  = {{Qwen Team}},
  journal = {arXiv preprint arXiv:2503.20215},
  year    = {2025}
}

@article{kong2024audioflamingo,
  title   = {{Audio Flamingo}: A Novel Audio Language Model with Few-Shot Learning and Dialogue Abilities},
  author  = {Kong, Zhifeng and Goel, Arushi and Badlani, Rohan and Ping, Wei and Valle, Rafael and Catanzaro, Bryan},
  journal = {arXiv preprint arXiv:2402.01831},
  year    = {2024}
}

@article{radford2023whisper,
  title     = {Robust Speech Recognition via Large-Scale Weak Supervision},
  author    = {Radford, Alec and Kim, Jong Wook and Xu, Tao and Brockman, Greg and McLeavey, Christine and Sutskever, Ilya},
  journal = {International Conference on Machine Learning (ICML)},
  year      = {2023}
}

@article{fang2025llamaomni2,
  title   = {{LLaMA-Omni2}: {LLM}-based Real-time Spoken Chatbot with Autoregressive Streaming Speech Synthesis},
  author  = {Fang, Qingkai and Zhou, Yan and Guo, Shoutao and Zhang, Shaolei and Feng, Yang},
  journal = {arXiv preprint arXiv:2505.02625},
  year    = {2025}
}

@article{gao2022paramformer,
  title   = {{Paraformer}: Fast and Accurate Parallel Transformer for Non-autoregressive End-to-End Speech Recognition},
  author  = {Gao, Zhifu and Zhang, Shiliang and McLoughlin, Ian and Yan, Zhijie},
  journal = {Proc. Interspeech},
  year    = {2022}
}

@article{li2025videochat,
  title={Videochat: Chat-centric video understanding},
  author={Li, KunChang and He, Yinan and Wang, Yi and Li, Yizhuo and Wang, Wenhai and Luo, Ping and Wang, Yali and Wang, Limin and Qiao, Yu},
  journal={Science China Information Sciences},
  volume={68},
  number={10},
  pages={200102},
  year={2025},
  publisher={Springer}
}

@inproceedings{chen2024videollm,
  title={Videollm-online: Online video large language model for streaming video},
  author={Chen, Joya and Lv, Zhaoyang and Wu, Shiwei and Lin, Kevin Qinghong and Song, Chenan and Gao, Difei and Liu, Jia-Wei and Gao, Ziteng and Mao, Dongxing and Shou, Mike Zheng},
  booktitle={Proceedings of the IEEE/CVF Conference on Computer Vision and Pattern Recognition},
  pages={18407--18418},
  year={2024}
}

@article{sakshi2024mmau,
  title={Mmau: A massive multi-task audio understanding and reasoning benchmark},
  author={Sakshi, Sakshi and Tyagi, Utkarsh and Kumar, Sonal and Seth, Ashish and Selvakumar, Ramaneswaran and Nieto, Oriol and Duraiswami, Ramani and Ghosh, Sreyan and Manocha, Dinesh},
  journal={arXiv preprint arXiv:2410.19168},
  year={2024}
}

@inproceedings{panayotov2015librispeech,
  title={Librispeech: an asr corpus based on public domain audio books},
  author={Panayotov, Vassil and Chen, Guoguo and Povey, Daniel and Khudanpur, Sanjeev},
  booktitle={2015 IEEE international conference on acoustics, speech and signal processing (ICASSP)},
  pages={5206--5210},
  year={2015},
  organization={IEEE}
}

@article{tian2025audiox,
  title={Audiox: Diffusion transformer for anything-to-audio generation},
  author={Tian, Zeyue and Jin, Yizhu and Liu, Zhaoyang and Yuan, Ruibin and Tan, Xu and Chen, Qifeng and Xue, Wei and Guo, Yike},
  journal={arXiv preprint arXiv:2503.10522},
  year={2025}
}

@inproceedings{gemmeke2017audio,
  title={Audio set: An ontology and human-labeled dataset for audio events},
  author={Gemmeke, Jort F and Ellis, Daniel PW and Freedman, Dylan and Jansen, Aren and Lawrence, Wade and Moore, R Channing and Plakal, Manoj and Ritter, Marvin},
  booktitle={2017 IEEE international conference on acoustics, speech and signal processing (ICASSP)},
  pages={776--780},
  year={2017},
  organization={IEEE}
}

@article{snyder2015musan,
  title={Musan: A music, speech, and noise corpus},
  author={Snyder, David and Chen, Guoguo and Povey, Daniel},
  journal={arXiv preprint arXiv:1510.08484},
  year={2015}
}

@article{wichern2019wham,
  title={Wham!: Extending speech separation to noisy environments},
  author={Wichern, Gordon and Antognini, Joe and Flynn, Michael and Zhu, Licheng Richard and McQuinn, Emmett and Crow, Dwight and Manilow, Ethan and Roux, Jonathan Le},
  journal={arXiv preprint arXiv:1907.01160},
  year={2019}
}

@article{chen2026voicebench,
  title={Voicebench: Benchmarking llm-based voice assistants},
  author={Chen, Yiming and Yue, Xianghu and Zhang, Chen and Gao, Xiaoxue and Tan, Robby T and Li, Haizhou},
  journal={Transactions of the Association for Computational Linguistics},
  volume={14},
  pages={378--398},
  year={2026},
  publisher={MIT Press 255 Main Street, 9th Floor, Cambridge, Massachusetts 02142, USA~…}
}

@article{ghosh2025audio,
  title={Audio flamingo 2: An audio-language model with long-audio understanding and expert reasoning abilities},
  author={Ghosh, Sreyan and Kong, Zhifeng and Kumar, Sonal and Sakshi, S and Kim, Jaehyeon and Ping, Wei and Valle, Rafael and Manocha, Dinesh and Catanzaro, Bryan},
  journal={arXiv preprint arXiv:2503.03983},
  year={2025}
}

@article{timcheck2023intel,
  title={The Intel neuromorphic DNS challenge},
  author={Timcheck, Jonathan and Shrestha, Sumit Bam and Ben Dayan Rubin, Daniel and Kupryjanow, Adam and Orchard, Garrick and Pindor, Lukasz and Shea, Timothy and Davies, Mike},
  journal={Neuromorphic Computing and Engineering},
  volume={3},
  number={3},
  pages={034005},
  year={2023},
  publisher={IOP Publishing}
}

@article{liu2025voxtral,
  title={Voxtral},
  author={Liu, Alexander H and Ehrenberg, Andy and Lo, Andy and Denoix, Cl{\'e}ment and Barreau, Corentin and Lample, Guillaume and Delignon, Jean-Malo and Chandu, Khyathi Raghavi and von Platen, Patrick and Muddireddy, Pavankumar Reddy and others},
  journal={arXiv preprint arXiv:2507.13264},
  year={2025}
}

@article{li2025baichuan,
  title={Baichuan-omni-1.5 technical report},
  author={Li, Yadong and Liu, Jun and Zhang, Tao and Chen, Song and Li, Tianpeng and Li, Zehuan and Liu, Lijun and Ming, Lingfeng and Dong, Guosheng and Pan, Da and others},
  journal={arXiv preprint arXiv:2501.15368},
  year={2025}
}

@article{abouelenin2025phi,
  title={Phi-4-mini technical report: Compact yet powerful multimodal language models via mixture-of-loras},
  author={Abouelenin, Abdelrahman and Ashfaq, Atabak and Atkinson, Adam and Awadalla, Hany and Bach, Nguyen and Bao, Jianmin and Benhaim, Alon and Cai, Martin and Chaudhary, Vishrav and Chen, Congcong and others},
  journal={arXiv preprint arXiv:2503.01743},
  year={2025}
}

@article{wang2024freeze,
  title={Freeze-omni: A smart and low latency speech-to-speech dialogue model with frozen llm},
  author={Wang, Xiong and Li, Yangze and Fu, Chaoyou and Shen, Yunhang and Xie, Lei and Li, Ke and Sun, Xing and Ma, Long},
  journal={arXiv preprint arXiv:2411.00774},
  year={2024}
}

@inproceedings{fang2025llama,
  title={LLaMA-omni 2: LLM-based real-time spoken chatbot with autoregressive streaming speech synthesis},
  author={Fang, Qingkai and Zhou, Yan and Guo, Shoutao and Zhang, Shaolei and Feng, Yang},
  booktitle={Proceedings of the 63rd Annual Meeting of the Association for Computational Linguistics (Volume 1: Long Papers)},
  pages={18617--18629},
  year={2025}
}

@inproceedings{wang2021covost,
  title={CoVoST 2 and massively multilingual speech translation.},
  author={Wang, Changhan and Wu, Anne and Gu, Jiatao and Pino, Juan},
  booktitle={Interspeech},
  volume={2021},
  pages={2247--2251},
  year={2021}
}

@article{goel2025audio,
  title={Audio flamingo 3: Advancing audio intelligence with fully open large audio language models},
  author={Goel, Arushi and Ghosh, Sreyan and Kim, Jaehyeon and Kumar, Sonal and Kong, Zhifeng and Lee, Sang-gil and Yang, Chao-Han Huck and Duraiswami, Ramani and Manocha, Dinesh and Valle, Rafael and others},
  journal={arXiv preprint arXiv:2507.08128},
  year={2025}
}

@article{sekoyan2025canary,
  title={Canary-1b-v2 \& parakeet-tdt-0.6 b-v3: Efficient and high-performance models for multilingual asr and ast},
  author={Sekoyan, Monica and Koluguri, Nithin Rao and Tadevosyan, Nune and Zelasko, Piotr and Bartley, Travis and Karpov, Nikolay and Balam, Jagadeesh and Ginsburg, Boris},
  journal={arXiv preprint arXiv:2509.14128},
  year={2025}
}

@article{xu2025fireredasr,
  title={Fireredasr: Open-source industrial-grade mandarin speech recognition models from encoder-decoder to llm integration},
  author={Xu, Kai-Tuo and Xie, Feng-Long and Tang, Xu and Hu, Yao},
  journal={arXiv preprint arXiv:2501.14350},
  year={2025}
}

@article{dubois2023alpacafarm,
  title={Alpacafarm: A simulation framework for methods that learn from human feedback},
  author={Dubois, Yann and Li, Chen Xuechen and Taori, Rohan and Zhang, Tianyi and Gulrajani, Ishaan and Ba, Jimmy and Guestrin, Carlos and Liang, Percy S and Hashimoto, Tatsunori B},
  journal={Advances in Neural Information Processing Systems},
  volume={36},
  pages={30039--30069},
  year={2023}
}

@inproceedings{faisal2021sd,
  title={SD-QA: Spoken dialectal question answering for the real world},
  author={Faisal, Fahim and Keshava, Sharlina and Alam, Md Mahfuz Ibn and Anastasopoulos, Antonios},
  booktitle={Findings of the Association for Computational Linguistics: EMNLP 2021},
  pages={3296--3315},
  year={2021}
}

@article{nachmani2023spoken,
  title={Spoken question answering and speech continuation using spectrogram-powered llm},
  author={Nachmani, Eliya and Levkovitch, Alon and Hirsch, Roy and Salazar, Julian and Asawaroengchai, Chulayuth and Mariooryad, Soroosh and Rivlin, Ehud and Skerry-Ryan, RJ and Ramanovich, Michelle Tadmor},
  journal={arXiv preprint arXiv:2305.15255},
  year={2023}
}

@inproceedings{berant2013semantic,
  title={Semantic parsing on freebase from question-answer pairs},
  author={Berant, Jonathan and Chou, Andrew and Frostig, Roy and Liang, Percy},
  booktitle={Proceedings of the 2013 conference on empirical methods in natural language processing},
  pages={1533--1544},
  year={2013}
}

@article{qwen3-asr,
  title={Qwen3-ASR Technical Report},
  author={Shi, Xian and Wang, Xiong and Guo, Zhifang and Wang, Yongqi and Zhang, Pei and Zhang, Xinyu and Guo, Zishan and Hao, Hongkun and Xi, Yu and Yang, Baosong and others},
  journal={arXiv preprint arXiv:2601.21337},
  year={2026}
}

@article{seamless,
  title={Seamless: Multilingual Expressive and Streaming Speech Translation},
  author={Barrault, Lo{\"\i}c and Chung, Yu-An and Meglioli, Mariano Coria and Dale, David and Dong, Ning and Duppenthaler, Mark and Duquenne, Paul-Ambroise and Ellis, Brian and Elsahar, Hady and Haaheim, Justin and others},
  journal={arXiv preprint arXiv:2312.05187},
  year={2023}
}

@inproceedings{lslm,
  title={Language model can listen while speaking},
  author={Ma, Ziyang and Song, Yakun and Du, Chenpeng and Cong, Jian and Chen, Zhuo and Wang, Yuping and Wang, Yuxuan and Chen, Xie},
  booktitle={Proceedings of the AAAI Conference on Artificial Intelligence},
  volume={39},
  number={23},
  pages={24831--24839},
  year={2025}
}

@article{miniomni2,
  title={Mini-omni2: Towards open-source gpt-4o with vision, speech and duplex capabilities},
  author={Xie, Zhifei and Wu, Changqiao},
  journal={arXiv preprint arXiv:2410.11190},
  year={2024}
}

@article{silent,
  title={The Silent Thought: Modeling Internal Cognition in Full-Duplex Spoken Dialogue Models via Latent Reasoning},
  author={Wu, Donghang and Zhang, Tianyu and Li, Yuxin and Liu, Hexin and Chen, Chen and Chng, Eng Siong and Bengio, Yoshua},
  journal={arXiv preprint arXiv:2603.17837},
  year={2026}
}

@article{chronological,
  title={Chronological thinking in full-duplex spoken dialogue language models},
  author={Wu, Donghang and Zhang, Haoyang and Chen, Chen and Zhang, Tianyu and Tian, Fei and Yang, Xuerui and Yu, Gang and Liu, Hexin and Hou, Nana and Hu, Yuchen and others},
  journal={arXiv preprint arXiv:2510.05150},
  year={2025}
}

@article{diffa2,
  title={DIFFA-2: A Practical Diffusion Large Language Model for General Audio Understanding},
  author={Zhou, Jiaming and Cheng, Xuxin and Zhao, Shiwan and Jia, Yuhang and Liu, Cao and Zeng, Ke and Cai, Xunliang and Qin, Yong},
  journal={arXiv preprint arXiv:2601.23161},
  year={2026}
}

@article{stepaudio2,
  title={Step-audio 2 technical report},
  author={Wu, Boyong and Yan, Chao and Hu, Chen and Yi, Cheng and Feng, Chengli and Tian, Fei and Shen, Feiyu and Yu, Gang and Zhang, Haoyang and Li, Jingbei and others},
  journal={arXiv preprint arXiv:2507.16632},
  year={2025}
}

@article{mmsu,
  title={Mmsu: A massive multi-task spoken language understanding and reasoning benchmark},
  author={Wang, Dingdong and Li, Junan and Wu, Jincenzi and Yang, Dongchao and Chen, Xueyuan and Zhang, Tianhua and Meng, Helen},
  journal={arXiv preprint arXiv:2506.04779},
  year={2025}
}

@article{seed-asr,
  title={Seed-asr: Understanding diverse speech and contexts with llm-based speech recognition},
  author={Bai, Ye and Chen, Jingping and Chen, Jitong and Chen, Wei and Chen, Zhuo and Ding, Chuang and Dong, Linhao and Dong, Qianqian and Du, Yujiao and Gao, Kepan and others},
  journal={arXiv preprint arXiv:2407.04675},
  year={2024}
}

@article{megaasr,
  title={Mega-ASR: Towards In-the-wild\^{} 2 Speech Recognition via Scaling up Real-world Acoustic Simulation},
  author={Xie, Zhifei and Pang, Kaiyu and Zhang, Haobin and Ye, Deheng and Hu, Xiaobin and Yan, Shuicheng and Miao, Chunyan},
  journal={arXiv preprint arXiv:2605.19833},
  year={2026}
}

@article{emotionthinker,
  title={EmotionThinker: Prosody-Aware Reinforcement Learning for Explainable Speech Emotion Reasoning},
  author={Wang, Dingdong and Liu, Shujie and Zhang, Tianhua and Chen, Youjun and Li, Jinyu and Meng, Helen},
  journal={arXiv preprint arXiv:2601.15668},
  year={2026}
}

@article{afcot,
  title={Audio Flamingo Sound-CoT Technical Report: Improving Chain-of-Thought Reasoning in Sound Understanding},
  author={Kong, Zhifeng and Goel, Arushi and Santos, Joao Felipe and Ghosh, Sreyan and Valle, Rafael and Ping, Wei and Catanzaro, Bryan},
  journal={arXiv preprint arXiv:2508.11818},
  year={2025}
}

@article{thinkwith,
  title={Thinking with sound: Audio chain-of-thought enables multimodal reasoning in large audio-language models},
  author={Xiong, Zhen and Cai, Yujun and Li, Zhecheng and Yuan, Junsong and Wang, Yiwei},
  journal={arXiv preprint arXiv:2509.21749},
  year={2025}
}

@article{audiocog,
  title={Audio-Cogito: Towards Deep Audio Reasoning in Large Audio Language Models},
  author={Li, Longhao and Chen, Hongjie and Li, Zehan and Hu, Qihan and Kang, Jian and Li, Jie and Xie, Lei and Li, Yongxiang},
  journal={arXiv preprint arXiv:2604.12527},
  year={2026}
}

@inproceedings{audio-reasoner,
  title={Audio-reasoner: Improving reasoning capability in large audio language models},
  author={Zhifei, Xie and Lin, Mingbao and Liu, Zihang and Wu, Pengcheng and Yan, Shuicheng and Miao, Chunyan},
  booktitle={Proceedings of the 2025 Conference on Empirical Methods in Natural Language Processing},
  pages={23840--23862},
  year={2025}
}

@article{proactive,
  title={Proactive agent research environment: Simulating active users to evaluate proactive assistants},
  author={Nathani, Deepak and Zhang, Cheng and Huan, Chang and Shan, Jiaming and Yang, Yinfei and Patel, Alkesh and Gan, Zhe and Wang, William Yang and Saxon, Michael and Wang, Xin Eric},
  journal={arXiv preprint arXiv:2604.00842},
  year={2026}
}

@article{proagent,
  title={ProAgent: Harnessing On-Demand Sensory Contexts for Proactive LLM Agent Systems},
  author={Yang, Bufang and Xu, Lilin and Zeng, Liekang and Guo, Yunqi and Jiang, Siyang and Lu, Wenrui and Liu, Kaiwei and Xiang, Hancheng and Jiang, Xiaofan and Xing, Guoliang and others},
  journal={arXiv preprint arXiv:2512.06721},
  year={2025}
}

@article{pask,
  title={PASK: Toward Intent-Aware Proactive Agents with Long-Term Memory},
  author={Xie, Zhifei and Hu, Zongzheng and Ye, Fangda and Zhang, Xin and Chai, Haobo and Liu, Zihang and Wu, Pengcheng and Zhang, Guibin and Liao, Yue and Hu, Xiaobin and others},
  journal={arXiv preprint arXiv:2604.08000},
  year={2026}
}

@article{needllm,
  title={Do Proactive Agents Really Need an LLM to Decide When to Wake and What to Anchor?},
  author={Liu, Xiaoze and Zhang, Ruowang and Abdi, Amir H and Galley, Michel and Chen, Zhikai and Xiong, Siheng and Wang, Xiaoqian and Gao, Jing},
  journal={arXiv preprint arXiv:2605.30152},
  year={2026}
}

@article{duplexsla,
  title={DuplexSLA: A Full-Duplex Spoken Language Model with Synchronized Speech, Language, and Action},
  author={Zhang, Haoyang and Chen, Jun and Wu, Donghang and Li, Yuxin and Zhang, Yuxin and Zhang, Xiangyu Tony and Liu, Che and Lin, Qingjian and Peng, Yizhou and Liu, Hexin and others},
  journal={arXiv preprint arXiv:2605.20755},
  year={2026}
}

\appendix

\section{Real-World Validation and Case Study}
\subsection{Real-World Validation}
\label{appendix:realworld}

To test transfer beyond synthetically composed streams, we collect approximately two hours of natural recordings from four deployment scenarios: \textit{Travel} (stations, airports, and hotel lobbies), \textit{Work} (meetings, typing, and notifications), \textit{Home} (domestic activity and staged safety events), and \textit{Commute} (walking, cycling, and in-vehicle audio). Recordings were captured at 16\,kHz using consumer smartphones and laptops, without TFJP or synthesis-time enhancements.

Trigger accuracy averages 58.9\%, compared with 62.0\% on a matched synthetic split. Performance declines most in \textit{Travel} and \textit{Commute}, where crowd and non-stationary noise increase ASR WER to approximately 7.9\% and 8.6\% and produce more proactive false positives. \textit{Work} remains closest to the synthetic baseline, while benign impulsive sounds slightly increase false positives at \textit{Home}. Average first-chunk latency remains within $\pm25$\,ms of the synthetic result across all scenarios, indicating that FIFO scheduling is robust to device and recording variation.

Internal diagnostics transfer consistently: per-chunk silence rates correlate at 0.91 with the synthetic split (Pearson correlation; 2\,s bins), ablating L35H14 reduces token match by 0.86 versus 0.88, and the GPT Layer~0 continuity ratio is 0.78 versus 0.80. These results indicate that the learned response policy transfers to natural recordings and is not driven primarily by synthetic concatenation cues.

\subsection{Case Study}

\begin{figure}[h]
    \centering
    \includegraphics[width=0.95\linewidth]{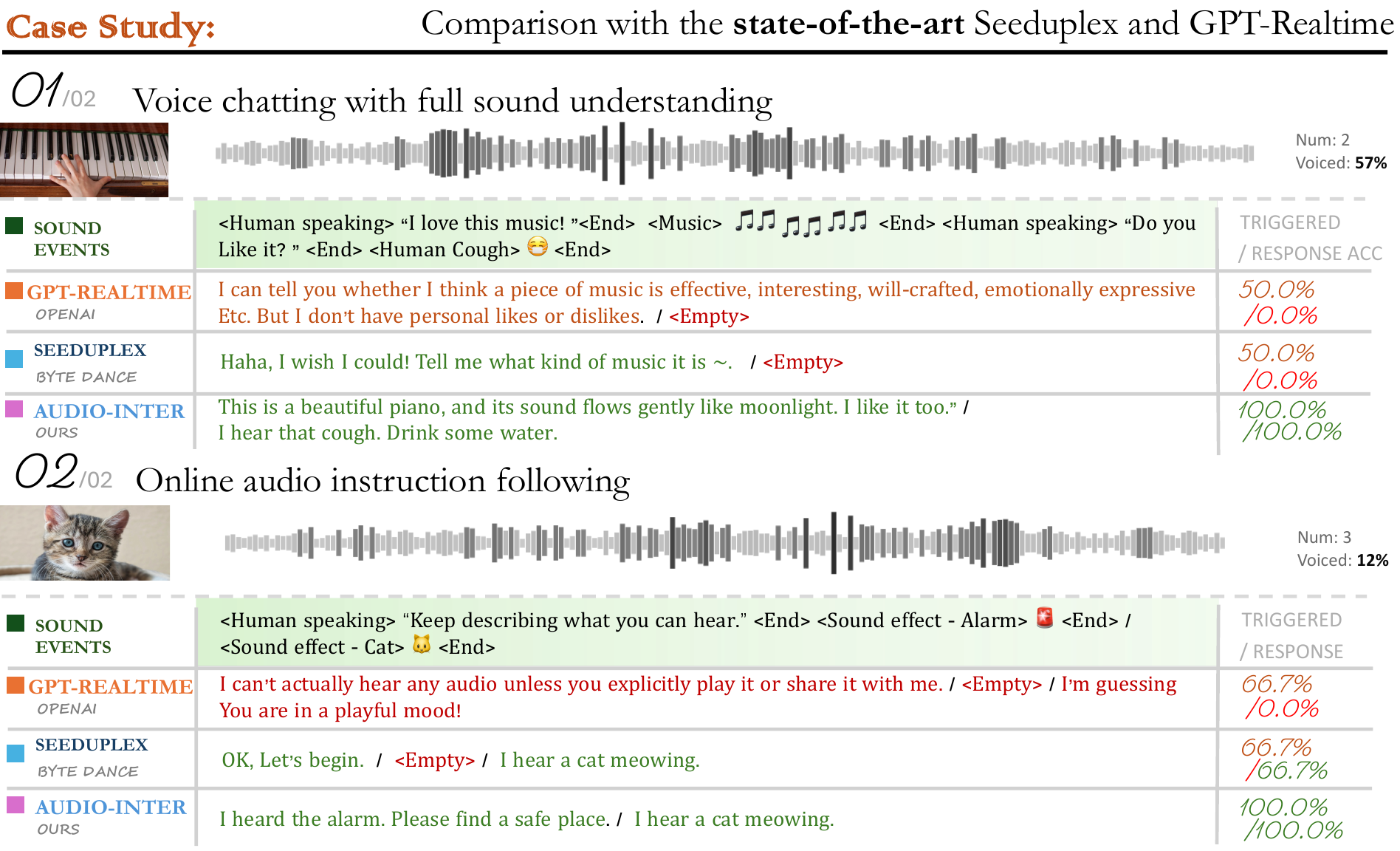}
    \caption{Qualitative streaming comparisons. In the cat example, baselines rely mainly on the transcribed word ``meow,'' whereas \textsc{Audio-Interaction} responds directly to the nonverbal acoustic cue and its surrounding context.}
    \label{fig:case_study}
\end{figure}

Figure\,\ref{fig:case_study} shows the full interaction loop. \textsc{Audio-Interaction} maintains the evolving scene, triggers on meaningful acoustic evidence, and responds directly to nonverbal events rather than relying on transcription shortcuts. This complements Proactive-Sound-Bench by showing how context continuity (C2) supports an appropriate intervention decision (C1).

\section{Method Details}
\label{appendix:method}

This section provides implementation details for the four components summarized in the main paper. Throughout, $c=400$\,ms is the streaming chunk size, and $f_{\text{enc}}$, $f_{\text{proj}}$, and $f_{\text{dec}}$ denote the audio encoder, adapter, and language model inherited from Qwen2.5-Omni-3B. Appendix\,\ref{appendix:hyperparams} reports all optimization hyperparameters.

\subsection{Streaming Data Construction}
\label{appendix:data-construction}

TFJP stabilizes clips before composition using six operators over a shared STFT representation. \textsc{silence\_cut} removes silent runs longer than $\tau$ using the 10th frame-energy percentile; \textsc{noise\_profile} estimates stationary noise from the lowest-energy 5\% of frames; and \textsc{denoise} applies spectral subtraction with $\gamma=1.0$. \textsc{core\_locate} selects the span maximizing normalized energy and spectral entropy, \textsc{boundary\_norm} aligns it to the nearest $\delta=c/2=200$\,ms boundary, and \textsc{spec\_smooth} applies a $\omega=20$\,ms Hann taper. We use $\tau=300$\,ms and $K=3$; fewer than 2\% of clips reach the iteration limit in Algorithm\,\ref{alg:tfjp}.

\begin{figure*}[!t]
    \centering
    \includegraphics[width=0.95\linewidth]{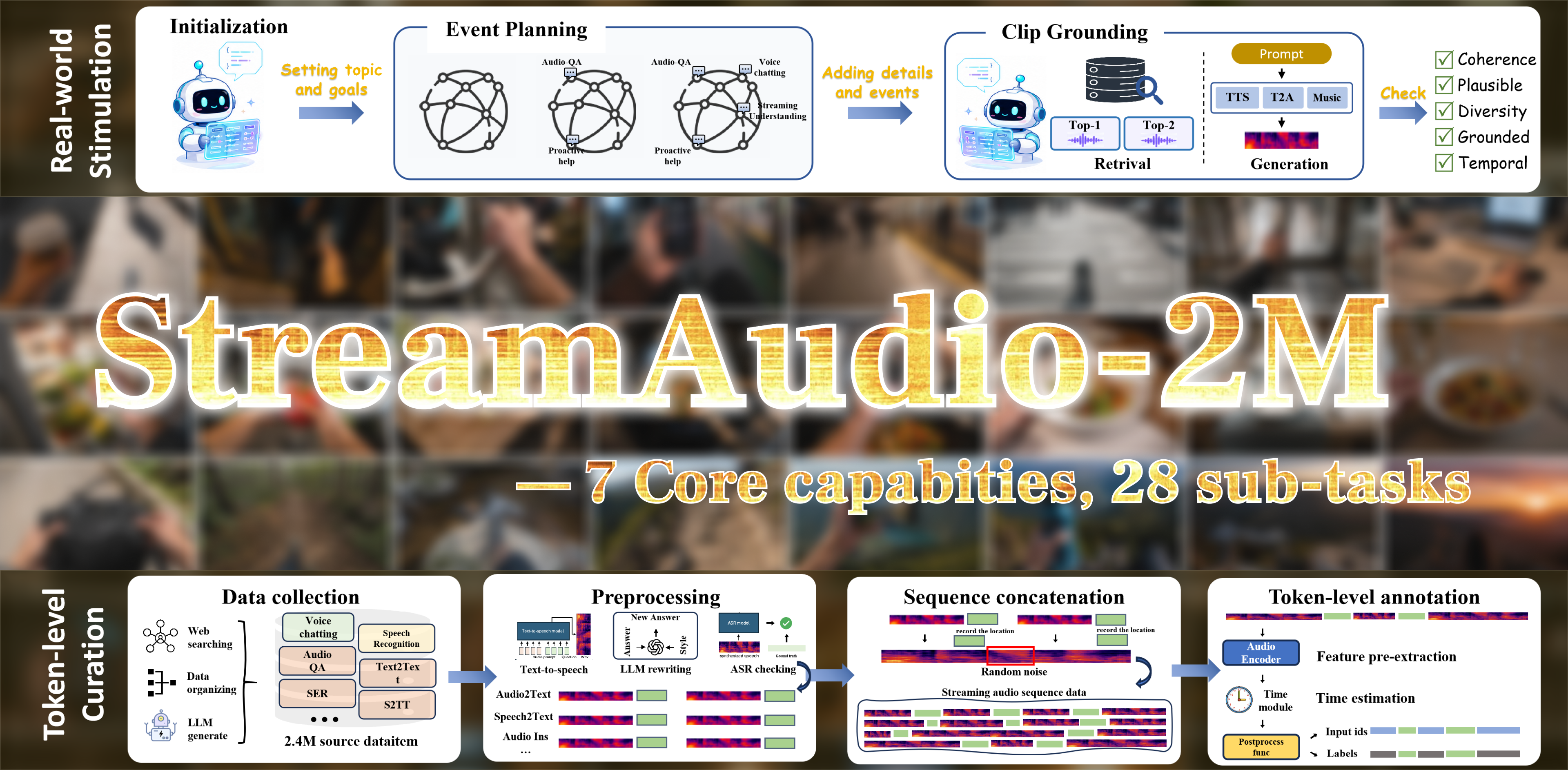}
    \caption{\textsc{StreamAudio-2M} pairs long-form simulated streams with token-level silence/response and text supervision across seven capabilities and 28 sub-tasks.}
    \label{fig:streamaudio_main}
\end{figure*}

Hierarchical event curation uses the prompts in Figures\,\ref{prompt:curation-p1}--\ref{prompt:curation-p2}. Stage~1 converts matched annotations into a coherent scenario containing 3--15 foreground, background, or ambient events. Stage~2 produces a retrieval query and generation fallback for each event. Stage~3 evaluates retrieved and synthesized clips using identity, cleanliness, duration fit, and continuity, returning \texttt{accept}, \texttt{reprocess}, or \texttt{reject}. All calls use JSON-mode decoding at temperature $0.7$.

\begin{algorithm}[t]
\caption{Time-Frequency Joint Preprocessing (TFJP).\label{alg:tfjp}}
\begin{algorithmic}[1]
\State \textbf{Input:} audio $x$, silence limit $\tau$, max iters $K$, smooth window $\omega$, align step $\delta$
\For{$k=1$ to $K$}
    \hfill {// S1--2: cut and norm.}
    \State $x \gets \texttt{silence\_cut}(x,\tau)$; $n \gets \texttt{noise\_profile}(x)$; $x \gets \texttt{denoise}(x,n)$
    \If{\texttt{stable}$(x,n)$} \textbf{break} \EndIf
\EndFor
\State $r \gets \texttt{core\_locate}(x)$ \hfill {// S3: localization}
\State $\tilde{x} \gets \texttt{boundary\_norm}(x,r,\tau,\delta)$ \hfill {// S4: trim.}
\State $\tilde{x} \gets \texttt{spec\_smooth}(\tilde{x},\omega)$
\State $x' \gets \texttt{silence\_cut}(\tilde{x},\tau)$ \hfill {// final check}
\If{\texttt{changed}$(x',\tilde{x})$} $x \gets x'$; \textbf{goto S1} \Else \Return $x'$ \EndIf
\end{algorithmic}
\end{algorithm}

\begin{algorithm}[h]
\caption{Streaming Sample Tokenization and Label Construction}
\label{alg:tokenize}
\begin{algorithmic}[1]
\Require instruction tokens $\mathcal{A}^{\text{ins}}$, audio chunks 
         $a_{1:T}$, response timeline 
         $\mathcal{R}=[(t_k, r_k)]_{k=1}^{K}$ sorted by $t_k$
\Ensure  token sequence $X$, streaming target $y^{\text{stream}}$, 
         LM target $y^{\text{LM}}$
\State $X, y^{\text{stream}}, y^{\text{LM}} \gets [\,],[\,],[\,]$
\State Append $\mathcal{A}^{\text{ins}}$ to $X$; extend labels with \textsc{Mask}
\State $k \gets 1$
\For{$t = 1$ to $T$}
    \State Append encoder features of $a_t$ to $X$; extend labels with \textsc{Mask}
    \If{$k \leq K \land t_k = t$}        \Comment{response triggers at chunk $t$}
        \State Append $\langle\texttt{response}\rangle$;\ 
               $y^{\text{stream}}\!{+\!=}\!\langle\texttt{response}\rangle$,\ 
               $y^{\text{LM}}\!{+\!=}\textsc{Mask}$
        \For{token $w$ in $r_k$}
            \State Append $w$;\ 
                   $y^{\text{stream}}\!{+\!=}\textsc{Mask}$,\ 
                   $y^{\text{LM}}\!{+\!=}w$
        \EndFor
        \State Append $\langle\texttt{eos}\rangle$;\ 
               $y^{\text{stream}}\!{+\!=}\textsc{Mask}$,\ 
               $y^{\text{LM}}\!{+\!=}\langle\texttt{eos}\rangle$
        \State $k \gets k+1$
    \Else                                \Comment{remain silent}
        \State Append $\langle\texttt{silent}\rangle$;\ 
               $y^{\text{stream}}\!{+\!=}\!\langle\texttt{silent}\rangle$,\ 
               $y^{\text{LM}}\!{+\!=}\textsc{Mask}$
    \EndIf
\EndFor
\State \Return $X, y^{\text{stream}}, y^{\text{LM}}$
\end{algorithmic}
\end{algorithm}

\subsection{Streaming Training}
\label{appendix:streaming-training}

A streaming sample carries two mutually exclusive supervision targets at
every position: $y^{\text{stream}}$ supervises one
$\langle\texttt{silent}\rangle$ or $\langle\texttt{response}\rangle$ token
per chunk; $y^{\text{LM}}$ supervises the text tokens following each
emitted $\langle\texttt{response}\rangle$. Audio-encoder positions and
the instruction prefix are masked from both. The construction is
formalized in Algorithm\,\ref{alg:tokenize}.

\begin{figure*}[t]
\centering
\includegraphics[width=0.95\linewidth]{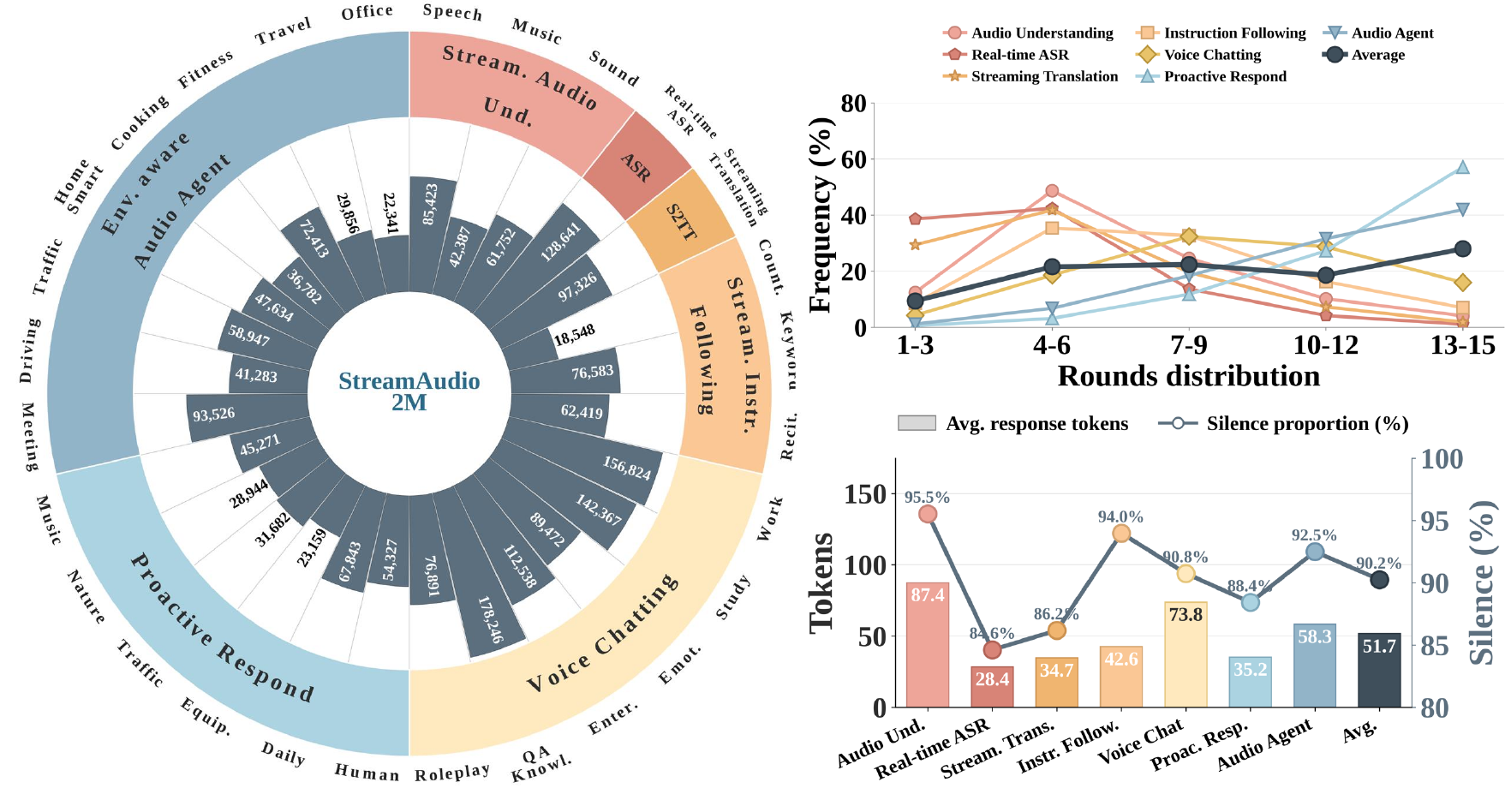}
\caption{Construction statistics for \textsc{StreamAudio-2M}. \textbf{(a)} Seven-capability taxonomy. \textbf{(b)} Round count, response length, and silence ratio by task.}
\label{fig:overview}
\end{figure*}

\begin{algorithm}[!h]
\caption{FIFO-Scheduled Asynchronous Streaming Inference}
\label{alg:fifo}
\begin{algorithmic}[1]
\Require audio stream $x_{1:\infty}$, encoder $f_{\text{enc}}$, decoder $f_{\text{dec}}$
\State \textbf{shared:} queue $\mathcal{Q}\!\gets\![\,]$;\ 
       last token $r^*\!\gets\!\langle\texttt{silent}\rangle$;\ 
       KV-cache $\mathcal{C}\!\gets\!\varnothing$
\State spawn \textsc{EncoderLoop} and \textsc{DecoderLoop} concurrently
\Statex
\Procedure{EncoderLoop}{}                  \Comment{producer; never blocks}
    \For{each arriving chunk $x_t$}
        \State $a_t \gets f_{\text{enc}}(x_t)$;\quad
               \textbf{atomic:} $\mathcal{Q}.\textsc{append}(a_t)$
    \EndFor
\EndProcedure
\Statex
\Procedure{DecoderLoop}{}                  \Comment{event-driven consumer}
    \Loop
        \If{$r^* \in \{\langle\texttt{silent}\rangle, \langle\texttt{eos}\rangle\}$}
            \State \textbf{wait until} $\mathcal{Q} \neq \varnothing$  
                   \Comment{idle if queue empty}
            \State \textbf{atomic:} $\mathcal{F}\!\gets\!\mathcal{Q}.\textsc{flush}()$;\quad
                   $\mathcal{C}\!\gets\!\textsc{Extend}(\mathcal{C},\mathcal{F})$
            \State $r^* \gets f_{\text{dec}}(\mathcal{C})$  
                   \Comment{emit one control token}
        \Else                              \Comment{mid-response}
            \State $r^* \gets f_{\text{dec}}(\mathcal{C})$  
                   \Comment{AR text step; queue untouched}
        \EndIf
        \State \textsc{Emit}($r^*$)
    \EndLoop
\EndProcedure
\end{algorithmic}
\end{algorithm}

Two dedicated prompts address insufficient long-range retention and false triggering (Figure\,\ref{prompt:supervision}). The history-review prompt generates follow-up questions whose answers depend on turns at least three rounds earlier. The silent-audio prompt evaluates non-speech segments using the four Proactive-Sound-Bench criteria; \texttt{borderline} clips are discarded rather than mislabeled. The dual-loss objective
$\mathcal{L}=\mathcal{L}_{\text{LM}}+\lambda\,\mathcal{L}_{\text{stream}}$
holds throughout the four-stage recipe; the recipe varies only the data
composition and trainable modules across stages: Stage~1 unfreezes the
LM head and the new-token embedding on offline single-turn data;
Stage~2 trains the adapter only; Stage~3 jointly trains adapter and LM
on the four core capabilities (ASR, S2TT, dialogue, audio understanding)
of StreamAudio-2M; Stage~4 fine-tunes on interleaved multi-turn
streams whose proactive insertions and history-review probes are
introduced as Bernoulli mix-ins during the composition pass of
\S\ref{appendix:curation}.

\subsection{Asynchronous FIFO Inference}
\label{appendix:fifo-inference}

The FIFO scheduler runs the encoder and decoder independently through queue $\mathcal{Q}$. The encoder consumes fixed-rate audio chunks and appends projected features without blocking on decoder state. The decoder is gated by its last emitted token $r^*$:
when $r^*\!\in\!\{\langle\texttt{silent}\rangle,
\langle\texttt{eos}\rangle\}$, the decoder is at an interruption point
and drains $\mathcal{Q}$ atomically into its KV-cache before emitting
one control token; when $r^*$ is a mid-response text token, the decoder
issues a pure autoregressive step against the existing KV-cache without
touching $\mathcal{Q}$. Flushing the queue at each interruption point keeps acoustic context aligned with wall-clock time after long responses and avoids stale silence decisions. This mechanism avoids decoder-induced listening stalls and keeps subsequent decisions synchronized with the incoming audio stream. Algorithm~\ref{alg:fifo} gives the complete schedule.

\subsection{Dataset Curation Pipeline}
\label{appendix:curation}

Text sources (MOSS, GammaCorpus, and instruction chats) are converted to speech in three steps. First, the prompt in Figure\,\ref{prompt:tts} removes visual formatting and expands numerals, abbreviations, and symbols. Second, CosyVoice renders each dialogue using one voice $v$ sampled from pool $\mathcal{V}$. Third, ASR verification rejects audio whose transcript exceeds $\tau_{\text{wer}}=0.10$ relative to the rewritten reference. We retry up to $R=2$ times, then discard the full dialogue to preserve multi-turn coherence.

Algorithm\,\ref{alg:compose} combines validated events with noise pool $\mathcal{N}=\textsc{MUSAN}\cup\textsc{WHAM!}\cup\textsc{DNS\text{-}Challenge}$. Foreground clips are concatenated with TFJP applied at each junction; background and ambient clips are mixed at random offsets using gains of $-6$ and $-12$\,dB. Event-like and ambient noise tracks span the full stream with crossfaded boundaries and SNRs sampled from $P_{\text{snr}}=\mathcal{U}(5,20)$\,dB, with ambient noise held 5\,dB quieter.

\begin{algorithm}[h]
\caption{Dual-Track Streaming Sequence Composition}
\label{alg:compose}
\begin{algorithmic}[1]
\Require ordered event list 
         $E\!=\![(w_i, \rho_i, d_i, r_i)]_{i=1}^{|E|}$ 
         (waveform, role, duration, response or $\varnothing$); 
         noise pool 
         $\mathcal{N}\!=\!\mathcal{N}_{\text{evt}}\uplus\mathcal{N}_{\text{amb}}$;
         chunk size $c$, fade window $\omega$, TFJP $\Phi$, 
         SNR distribution $P_{\text{snr}}$
\Ensure  stream waveform $y$, response timeline $\mathcal{T}$
\State $y_{\text{main}}\!\gets\!\varnothing$;\quad $\mathcal{T}\!\gets\![\,]$
\For{$i = 1$ to $|E|$}
    \State $w_i \gets \Phi(w_i)$  
           \Comment{re-apply TFJP at clip boundary}
    \If{$\rho_i = \texttt{foreground}$}
        \State $\textit{offset} \gets \textsc{Length}(y_{\text{main}})$;\quad
               $y_{\text{main}} \gets \textsc{Concat}(y_{\text{main}},\textsc{Fade}(w_i,\omega))$
        \If{$r_i \neq \varnothing$}
            \State $\mathcal{T}.\textsc{append}\big(\big(\lceil(\textit{offset}+d_i)/c\rceil,\ r_i\big)\big)$
        \EndIf
    \Else                                  \Comment{$\rho_i \in \{\texttt{bg},\texttt{amb}\}$}
        \State $\textsc{MixIn}(y_{\text{main}},\,w_i,\,
               \text{rand offset},\,\textsc{RoleGain}(\rho_i))$
    \EndIf
\EndFor
\State $D \gets \textsc{Length}(y_{\text{main}})$
\State $y^{(1)}\!\gets\!\textsc{TileCrossfade}(\textsc{Sample}(\mathcal{N}_{\text{evt}}),\,D)$;\quad 
       $y^{(2)}\!\gets\!\textsc{TileCrossfade}(\textsc{Sample}(\mathcal{N}_{\text{amb}}),\,D)$
\State $\sigma_1 \sim P_{\text{snr}}$;\quad $\sigma_2 \sim P_{\text{snr}}+5\,\text{dB}$
       \Comment{ambient held quieter}
\State $y \gets y_{\text{main}} 
              + \textsc{Scale}(y^{(1)},\sigma_1) 
              + \textsc{Scale}(y^{(2)},\sigma_2)$
\State \Return $y,\,\mathcal{T}$
\end{algorithmic}
\end{algorithm}

The resulting waveform $y$ and response timeline $\mathcal{T}$ feed Algorithm\,\ref{alg:tokenize}, which creates the $\langle X,y^{\text{stream}},y^{\text{LM}}\rangle$ tuple over 400\,ms chunks. All seven task families use this procedure; they differ only in response timing. ASR responds per chunk, dialogue at user-turn boundaries, and proactive assistance only at designated events.

\section{StreamAudio-2M Dataset Sources}
\label{app:dataset}

\textsc{StreamAudio-2M} draws from public corpora selected for complementary roles in streaming interaction. We favor established sources for reproducibility, then transform them through speech synthesis, chunking, event composition, and environmental mixing. Table\,\ref{tab:app:sources} summarizes each source and its pre-composition contribution.

\begin{table}[t]
\centering
\small
\setlength{\tabcolsep}{3.5pt}
\renewcommand{\arraystretch}{1.08}
\begin{tabularx}{\columnwidth}{@{}>{\raggedright\arraybackslash}X r r@{}}
\toprule
\textbf{Task family} & \textbf{\#Items} & \textbf{Share} \\
\midrule
Voice Chatting & 539k & 23.1\% \\
\multicolumn{3}{@{}p{\columnwidth}@{}}{\textit{Sources:} MOSS; GammaCorpus-Fact-QA} \\
Streaming Instr. Following & 487k & 20.8\% \\
\multicolumn{3}{@{}p{\columnwidth}@{}}{\textit{Sources:} UltraChat; Magpie-Pro; BellGroup; COIG-CQIA; DU-QA; Web-QA} \\
Streaming Audio Understanding & 382k & 16.4\% \\
\multicolumn{3}{@{}p{\columnwidth}@{}}{\textit{Sources:} AudioSet (Open/Choice); FMA (Open/Choice)} \\
Streaming Translation & 357k & 15.3\% \\
\multicolumn{3}{@{}p{\columnwidth}@{}}{\textit{Sources:} CoVoST 2 (En$\leftrightarrow$CN); AISHELL} \\
Real-time ASR & 270k & 11.6\% \\
\multicolumn{3}{@{}p{\columnwidth}@{}}{\textit{Sources:} CommonVoice; GigaSpeech; LibriSpeech; VoxPopuli} \\
Proactive Response & 171k & 7.3\% \\
\multicolumn{3}{@{}p{\columnwidth}@{}}{\textit{Sources:} AudioSet (events); AudioX; ElevenLabs} \\
Environmental Audio Agent & 130k & 5.5\% \\
\multicolumn{3}{@{}p{\columnwidth}@{}}{\textit{Sources:} MOSS; AudioSet (Desc./Open); WHAM!; DNS; MUSAN} \\
\midrule
\multicolumn{3}{@{}c@{}}{\textbf{2.34M}\,items; \textbf{7.49M}\,rounds; \textbf{66.7K}\,hrs} \\
\bottomrule
\end{tabularx}
\caption{Sources for \textsc{StreamAudio-2M}.}
\label{tab:overview_sources}
\end{table}

\begin{table*}[t]
\centering
\small
\setlength{\tabcolsep}{4pt}
\renewcommand{\arraystretch}{1.05}
\begin{tabularx}{\linewidth}{@{}l l X r r@{}}
\toprule
\textbf{Source} & \textbf{Family} & \textbf{Role in StreamAudio-2M} & \textbf{Items} & \textbf{Hours} \\
\midrule
CommonVoice           & Speech         & Streaming ASR supervision (multilingual)          & 62{,}354  & 120 \\
GigaSpeech            & Speech         & Streaming ASR supervision (in-the-wild)           & 86{,}740  & 170 \\
LibriSpeech           & Speech         & Streaming ASR supervision (read speech)           & 81{,}647  & 160 \\
VoxPopuli             & Speech         & Streaming ASR supervision (parliamentary)         & 39{,}746  & 80  \\
\midrule
CoVoST 2 (En$\to$Zh)  & Speech         & Speech translation \& simultaneous interpretation & 198{,}942 & 390 \\
CoVoST 2 (Zh$\to$En)  & Speech         & Speech translation \& simultaneous interpretation & 16{,}826  & 35  \\
AISHELL               & Speech         & Mandarin ASR / translation supervision            & 141{,}246 & 280 \\
\midrule
FMA (Open)            & Audio          & Open-ended music understanding prompts            & 33{,}154  & \multirow{2}{*}{150} \\
FMA (Choice)          & Audio          & Multiple-choice music understanding prompts       & 42{,}347  &     \\
AudioSet (Open)       & Audio          & Open-ended audio-QA grounding events              & 171{,}030 & \multirow{3}{*}{820} \\
AudioSet (Choice)     & Audio          & Multiple-choice audio-QA reasoning prompts        & 135{,}753 &     \\
AudioSet (Description)& Audio          & Audio captioning \& scene description             & 99{,}946  &     \\
\midrule
MOSS                  & Speech         & Spoken-dialogue supervision (TTS-rendered)        & 392{,}198 & 4{,}900 \\
GammaCorpus-Fact-QA   & Speech         & Factual spoken-QA supervision (TTS-rendered)      & 147{,}253 & 1{,}840 \\
\midrule
AudioSet (events)     & Acoustic event & Real foreground events for streams                & 27{,}491  & \multirow{3}{*}{160} \\
AudioX                & Acoustic event & Synthesized rare-event clips                      & 94{,}503  &     \\
ElevenLabs            & Acoustic event & Synthesized targeted sound effects                & 48{,}927  &     \\
\midrule
MUSAN                 & Noise          & Music, speech and ambient background              & 1{,}896   & \multirow{3}{*}{620} \\
WHAM!                 & Noise          & Real-world reverberant scenes                     & 13{,}425  &     \\
DNS Challenge         & Noise          & Diverse environmental conditions                  & 14{,}328  &     \\
\midrule
UltraChat             & Auxiliary      & Text-only instruction following (multi-turn)      & 156{,}732 & --  \\
Magpie-Pro            & Auxiliary      & Text-only instruction following (self-aligned)    & 167{,}324 & --  \\
DU-QA                 & Auxiliary      & Text-only domain-understanding QA                 & 14{,}308  & --  \\
COIG-CQIA             & Auxiliary      & Chinese instruction following                     & 34{,}274  & --  \\
Web-QA                & Auxiliary      & Open-domain web question answering                & 5{,}892   & --  \\
BellGroup             & Auxiliary      & Chinese conversational instructions               & 108{,}173 & --  \\
\bottomrule
\end{tabularx}
\caption{Source corpora used to construct \textsc{StreamAudio-2M}. Items and hours denote pre-composition instance counts and raw audio duration; text-only auxiliary sources are marked ``--'' under Hours.}
\label{tab:app:sources}
\end{table*}

\paragraph{Speech-centric sources.}
Speech-centric sources preserve four core capabilities: spoken dialogue, ASR, speech translation, and audio question answering. \textbf{MOSS} contributes 392k multi-turn dialogues, rendered into 4{,}900 hours of speech using multi-voice CosyVoice. \textbf{LibriSpeech}~\citep{panayotov2015librispeech} is re-segmented into 400\,ms chunks so recognition targets arrive during listening rather than at utterance end. \textbf{CoVoST 2}~\citep{wang2021covost} contributes 216k English--Chinese pairs used both as offline translation examples and as interleaved timelines for simultaneous interpretation.

\paragraph{Acoustic event sources.}
Streaming interaction requires both common foreground events and rare, context-dependent cues. \textbf{AudioSet} supplies real recordings sampled broadly across its ontology. For sparsely covered safety-critical sounds, \textbf{AudioX}~\citep{tian2025audiox} and \textbf{ElevenLabs} generate candidates that pass the same verification pipeline as retrieved clips. Together, real and synthetic sources provide 171k events covering the Proactive-Sound-Bench taxonomy.

\begin{table*}[t]
  \centering
  \small
  \setlength{\tabcolsep}{6pt}
  \renewcommand{\arraystretch}{1.3}
  \begin{tabular}{@{}
      >{\raggedright\arraybackslash}p{3.05cm}
      >{\raggedright\arraybackslash}p{2.35cm}
      p{0.545\linewidth} @{}}
    \toprule
    \textbf{Meso subdomain} & \textbf{Macro domain} & \textbf{Definition} \\
    \midrule
    Body Movements      & Human & Sounds associated with physical activity, falls, and injury. \\
    Physiological States & Human & Normal bodily functions and acute physiological distress. \\
    Emotional Expression & Human & Affective vocalizations and expressive nonverbal signals. \\
    Collective Ambience  & Human & Crowd-dominated environments and group activity. \\
    \midrule
    Personal Care       & Daily Living & Self-care activities in domestic environments. \\
    Daily Affairs       & Daily Living & Routine interactions with furniture, handheld objects, and indoor surfaces. \\
    Housekeeping        & Daily Living & Cleaning, tidying, and repetitive maintenance activities. \\
    \midrule
    Household Equipment & Equipment & Normal and abnormal operation of household appliances and systems. \\
    Industrial Tools    & Equipment & Powered tools, industrial machinery, and high-energy mechanical transients. \\
    \midrule
    Vehicle             & Traffic & Mechanical and warning sounds produced by individual vehicles. \\
    Traffic             & Traffic & Road ambience and intermittent urban warning signals. \\
    Mass Transit        & Traffic & Rail, bus, and heavy-vehicle environments with large mechanical resonances. \\
    \midrule
    Meteorological Events & Environment & Weather-driven airborne and precipitation sounds, from calm ambience to severe storms. \\
    Geological Hazards  & Environment & Terrain movement, rockfalls, and other indicators of geological instability. \\
    Ecological Context  & Environment & Outdoor biological cues from animals and ecosystems. \\
    Social Places       & Environment & Ambient soundscapes in occupied public and social spaces. \\
    \midrule
    Instrumental Music  & Music       & Instrumental performance and damage-related acoustic anomalies. \\
    \bottomrule
  \end{tabular}
  \caption{Meso-level category definitions for Proactive-Sound-Bench.}
  \label{tab_meso_definitions}
\end{table*}

\paragraph{Noise sources.}
Every long-form stream receives two background tracks so the model learns to ignore persistent non-foreground audio. \textbf{MUSAN}~\citep{snyder2015musan} provides music, ambient, and speech-babble noise; \textbf{WHAM!}~\citep{wichern2019wham} provides urban and reverberant scenes; and \textbf{DNS Challenge}~\citep{timcheck2023intel} broadens environmental coverage. These sources contribute 620 hours mixed at controlled SNRs rather than treated as response-bearing events.

\section{Proactive-Sound-Bench}
\label{app:bench}

\subsection{Task Definition}
\label{app:bench:task}

\begin{figure}[!h]
    \centering
    \includegraphics[width=1\linewidth]{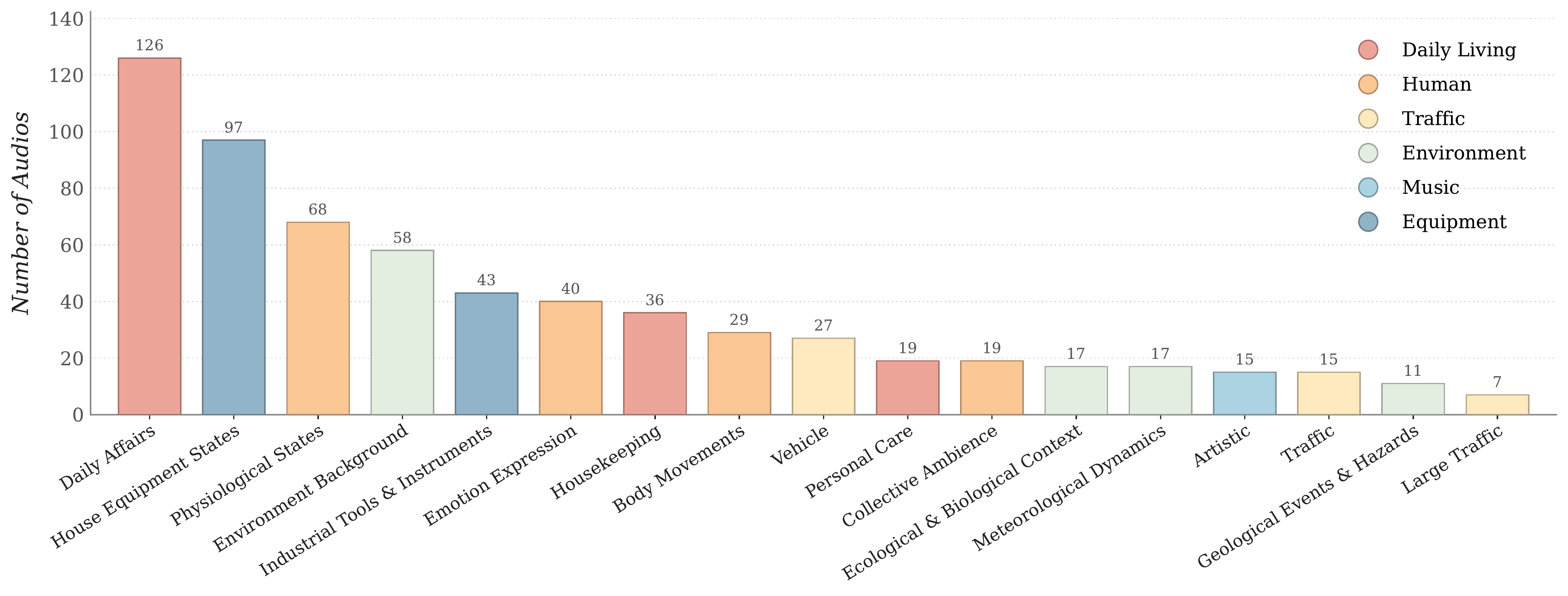}
    \caption{Distribution of Proactive-Sound-Bench examples across six macro domains and 17 meso-level subdomains.}
    \label{fig:proactive_distribution}
\end{figure}

Proactive-Sound-Bench evaluates two coupled abilities from an audio stream $x$: \textbf{(i)} deciding whether intervention is warranted and \textbf{(ii)} generating an appropriate response when triggered. The model should respond to acute physiological distress, severe weather, potential equipment damage, or hazardous environmental signals, while remaining silent for normal bodily sounds, routine device operation, and other benign events. Triggered responses should provide actionable warnings, suggestions, or first-aid guidance rather than generic acknowledgments.

Unlike sound event detection, which predicts event labels or temporal boundaries, and audio captioning, which neutrally describes a scene, Proactive-Sound-Bench evaluates both \emph{whether to respond} and \emph{what to say}. Semantic matching against multiple references accommodates valid response diversity. The task therefore tests whether acoustic understanding can be converted into context-sensitive interaction decisions.

\subsection{Categories and Coverage}
\label{app:bench:categories}

\paragraph{Taxonomy rationale.}
The taxonomy covers six everyday acoustic domains. \textbf{Human} contains physiological, emotional, movement, and crowd cues; \textbf{Daily Living} covers object handling and domestic routines; \textbf{Equipment}, \textbf{Traffic}, and \textbf{Nature \& Environment} capture engineered and outdoor hazards; and \textbf{Music} covers normal performance and damage-related anomalies. This organization separates acoustically similar events by their interaction context and expected response policy.

\section{Experimental Details}
\label{appendix:hyperparams}

Table\,\ref{tab:hyperparams} reports the method, data, optimization, and hardware settings for all four training stages. Method constants follow the main-paper ablations, while data parameters follow Appendix\,\ref{appendix:curation}. Stage-specific learning rates and step budgets reflect the amount of data and number of trainable parameters. Training uses bf16 mixed precision, gradient checkpointing, and DeepSpeed ZeRO-2 on $32\!\times\!\textsc{NVIDIA H100}$ 80\,GB GPUs.

\begin{table*}[t]
\centering
\small
\setlength{\tabcolsep}{4pt}
\renewcommand{\arraystretch}{1.05}
\begin{tabular}{c|l|cccc}
\toprule
\textbf{Configurations} & \textbf{Parameters} & \multicolumn{4}{c}{\textbf{Values}} \\
\cmidrule(lr){3-6}
 &  & \textbf{Stage 1} & \textbf{Stage 2} & \textbf{Stage 3} & \textbf{Stage 4} \\
\midrule
\multirow{5}{*}{Streaming}
  & chunk size $c$                   & \multicolumn{4}{c}{$400$\,ms} \\
  & fade window $\omega$             & \multicolumn{4}{c}{$20$\,ms} \\
  & half-chunk align $\delta$        & \multicolumn{4}{c}{$200$\,ms} \\
  & dual-loss weight $\lambda$       & \multicolumn{4}{c}{$1.0$} \\
  & max stream length $L_{\max}$     & \multicolumn{4}{c}{$60$ chunks ($24$\,s)} \\
\midrule
\multirow{7}{*}{Data}
  & WER threshold $\tau_{\text{wer}}$ & \multicolumn{4}{c}{$0.10$} \\
  & ASR retries $R$                  & \multicolumn{4}{c}{$2$} \\
  & SNR distribution $P_{\text{snr}}$ & \multicolumn{4}{c}{$\mathcal{U}(5,\,20)$\,dB} \\
  & role gain (fg / bg / amb)        & \multicolumn{4}{c}{$0$ / $-6$ / $-12$\,dB} \\
  & history-review prob $p_{\text{hr}}$  & --- & --- & --- & $0.30$ \\
  & silent mix prob $p_{\text{sil}}$     & --- & --- & --- & $0.40$ \\
  & proactive mix prob $p_{\text{pro}}$  & --- & --- & --- & $0.30$ \\
\midrule
\multirow{11}{*}{Training}
  & trainable modules         & LM head + emb.       & adapter              & adapter + LM         & adapter + LM \\
  & batch size (per GPU)      & $8$                  & $8$                  & $4$                  & $2$ \\
  & gradient accum.\ steps    & $2$                  & $4$                  & $8$                  & $16$ \\
  & effective batch size      & $512$                & $1024$               & $1024$               & $1024$ \\
  & learning rate             & $1\!\times\!10^{-4}$ & $1\!\times\!10^{-4}$ & $5\!\times\!10^{-5}$ & $1\!\times\!10^{-5}$ \\
  & training steps            & $5$\,k               & $20$\,k              & $80$\,k              & $15$\,k \\
  & warmup ratio              & $0.03$               & $0.03$               & $0.03$               & $0.03$ \\
  & optimizer                 & \multicolumn{4}{c}{AdamW ($\beta_1\!=\!0.9$, $\beta_2\!=\!0.95$, $\varepsilon\!=\!10^{-8}$)} \\
  & scheduler                 & \multicolumn{4}{c}{Cosine decay with linear warmup} \\
  & weight decay              & \multicolumn{4}{c}{$0.01$} \\
  & max grad norm             & \multicolumn{4}{c}{$1.0$} \\
\midrule
\multirow{3}{*}{Hardware}
  & GPUs                      & \multicolumn{4}{c}{$32\!\times\!\textsc{NVIDIA H100}$ $80$\,GB} \\
  & precision \& sharding     & \multicolumn{4}{c}{bf16 mixed precision, DeepSpeed ZeRO-2} \\
  & total wall-clock time     & \multicolumn{4}{c}{$\sim 10$ days} \\
\bottomrule
\end{tabular}
\caption{Method, data, training, and hardware configurations for \textsc{Audio-Interaction}.}
\label{tab:hyperparams}
\end{table*}

\section{Extended Related Work}

\paragraph{Streaming Audio Models.}
Streaming audio models generally specialize in a single function, including speech recognition~\citep{gao2022paramformer}, speech translation~\citep{seamless}, or full-duplex dialogue~\citep{lslm, miniomni2, silent, chronological}. DuplexSLA~\citep{duplexsla} further incorporates actions. Like Moshi~\citep{defossez2024moshi}, \textsc{Audio-Interaction} processes fixed-size chunks and decides when to intervene. Its policy is broader, however, because it must jointly reason over speech, environmental sounds, paralinguistic cues, long-range context, and explicit instructions.

\paragraph{Audio Large Models.}
Audio language models increasingly unify speech, sound, and music understanding~\citep{chu2024qwen2, qwen25omni2025, diffa2, stepaudio2}. Their capabilities include general audio understanding~\citep{sakshi2024mmau}, spoken dialogue~\citep{mmsu}, speech recognition~\citep{xu2025fireredasr, seed-asr, qwen3-asr, megaasr}, emotion understanding~\citep{emotionthinker}, and audio reasoning~\citep{afcot, thinkwith, audiocog, audio-reasoner}. Most nevertheless retain an offline interface in which a complete clip precedes generation. Our work extends this generality to continuous, instruction-conditioned interaction.

\paragraph{Streaming AI Systems.}
Related streaming systems also appear in online video understanding~\citep{chen2024videollm, li2025videochat} and proactive agents~\citep{proactive, proagent, pask, needllm}. Video models process incoming frames continuously, whereas proactive agents often cascade perception, textualization, and decision modules. \textsc{Audio-Interaction} instead realizes continuous perception, intervention timing, and response generation within one audio-native model.

\section{Error Analysis}
\label{app:analysis}

\label{app:analysis:err:breakdown}
\begin{itemize}
    \item \textbf{LibriSpeech (ASR).}
    Among 98 non-empty, non-crash errors, local token deviations---phonetic or orthographic substitutions and minor insertions/deletions---account for 60.2\%. Rare-word and long-utterance degradation contributes 21.4\%, primarily through named-entity errors and structural failures in complex sentences. Function-word bias accounts for 14.3\%, and decoding loops for 4.1\%.
  
    \item \textbf{CoVoST2 (Speech-to-Text Translation).}
    For En$\rightarrow$Zh examples with BLEU below 20, semantic hallucinations dominate (82\%); the remaining 18\% contain omissions, untranslated English fragments, or malformed Chinese. For the lowest-scoring Zh$\rightarrow$En examples, 75.5\% are off-topic or hallucinated, likely reflecting recognition or alignment failures, while 24.5\% contain omissions or uncontrolled paraphrases that preserve only part of the meaning.

    \item \textbf{MMAU (Audio Understanding).}
    Generation collapse produces unparseable outputs in approximately 20\% of errors. The remaining 80\% are recognition or reasoning failures, including confusion between similar sound sources, incorrect speaker-attribute judgments, and wrong category selection despite partially correct reasoning.

    \item \textbf{SpokenQA (Llama Questions and Web Questions).}
    After removing 35 empty predictions and formatting-related false errors, Llama Questions contains 37 genuine errors: factual hallucinations (56.8\%), irrelevant or overly general responses (27.0\%), and temporal or quantitative mistakes (16.2\%). Web Questions shows a similar pattern, with approximately 71\% factual hallucinations and about 15\% each for irrelevant responses and temporal or numerical errors.

    \item \textbf{VoiceBench (AlpacaEval-full and SD-QA).}
    Low-scoring AlpacaEval examples consist of hallucinations (53.5\%) and irrelevant responses or inappropriate refusals (46.4\%). SD-QA errors are primarily factual hallucinations (63\%), followed by irrelevant or miscomprehending responses (24\%) and over-refusals on benign questions (13\%).

    \item \textbf{Proactive-Sound-Bench.}
    False positives account for 59.8\% of errors and mainly arise from overreacting to benign sounds such as tearing paper, appliance noise, drinking, or sighing. False negatives account for 40.2\% and cluster around safety-critical traffic alarms and natural hazards.

\end{itemize}

\begin{figure}[h]
\centering
\begin{tcolorbox}[
  colback=white, colbacktitle=white, colframe=black,
  coltitle=black,
  title=\textbf{Prompt Template (Hierarchical Event Curation, Part 1)},
  fonttitle=\bfseries,
  boxrule=0.5pt, arc=2pt,
  fontupper=\small,
  left=8pt, right=8pt, top=4pt, bottom=4pt
]
\textbf{Stage 1 --- Scenario Planning}\\[1pt]
\textbf{System:} You are an audio scenario designer. Given a small bag of
unrelated audio annotations drawn from heterogeneous corpora, compose
a single coherent real-world scene that could plausibly contain all of
them. The scene must satisfy three properties: \textbf{(a) temporal
ordering} --- events flow in a physically plausible sequence;
\textbf{(b) acoustic compatibility} --- outdoor cues should not co-occur
with closed-room cues without an explicit transition (e.g., a door
opening); \textbf{(c) role consistency} --- foreground events must not
collide with incompatible background ambiences (a quiet library reading
cannot share an ambient layer with crowd cheering). Each sub-event takes
one role: \textit{foreground} (focal event the listener attends to),
\textit{background} (recurrent non-focal sound, mixed lower), or
\textit{ambient} (continuous texture spanning the whole scene). If two
annotations are mutually exclusive (e.g., ``indoor cooking'' and
``highway driving''), output two scenes rather than forcing one.\\[2pt]
\textbf{User:} \texttt{annotations}: \verb|{annotations}|.\\[2pt]
\textbf{Required output (JSON):}
\begin{itemize}
  \item \texttt{scenario}: one-sentence description, including location
        and time of day.
  \item \texttt{sub\_events}: ordered list of 3--15 entries with
        \texttt{name}, \texttt{role} (\texttt{fg}/\texttt{bg}/\texttt{amb}),
        duration, and brief description; at least one ambient slot must
        cover the full duration.
  \item \texttt{constraints}: ordering / mutual-exclusion rules that
        must hold downstream (e.g., ``doorbell precedes footsteps'';
        ``music and TV cannot overlap'').
\end{itemize}
Return only the JSON object, no surrounding prose.

\rule{\linewidth}{0.4pt}\\[2pt]

\textbf{Stage 2 --- Event Refinement}\\[1pt]
\textbf{System:} For each sub-event, produce a concrete query string
that an audio retrieval engine can match against AudioSet-style ontology
tags and free-text captions. Queries must be specific enough to
discriminate near-confusables --- write ``ceramic mug placed on wood''
rather than ``object on table'', and ``metal spoon stirring in glass''
rather than ``stirring sound''. Also produce a fallback caption that
describes the same event with enough acoustic detail (material, surface,
intensity, duration) to drive a generative SFX model when retrieval
fails.\\[2pt]
\textbf{User:} \texttt{scenario}: \verb|{scenario}|;\quad
\texttt{sub\_events}: \verb|{sub_events}|.\\[2pt]
\textbf{Required output (JSON list, aligned to input order):}
\begin{itemize}
  \item \texttt{query}: 4--12-word retrieval query.
  \item \texttt{fallback\_caption}: 1--2-sentence caption used by the
        audio generator if retrieval fails; must include material,
        surface and intensity cues.
  \item \texttt{accept\_tags}: small set of AudioSet ontology tags whose
        presence is sufficient evidence of a match.
\end{itemize}
\end{tcolorbox}
\caption{Prompt template for hierarchical event curation, Part 1:
scenario planning followed by event refinement. Both calls run in
JSON-mode decoding at temperature $0.7$.}
\label{prompt:curation-p1}
\end{figure}

\begin{figure}[h]
\centering
\begin{tcolorbox}[
  colback=white, colbacktitle=white, colframe=black,
  coltitle=black,
  title=\textbf{Prompt Template (Hierarchical Event Curation, Part 2)},
  fonttitle=\bfseries,
  boxrule=0.5pt, arc=2pt,
  fontupper=\small,
  left=8pt, right=8pt, top=4pt, bottom=4pt
]
\textbf{Stage 3 --- Clip Grounding Verification}\\[1pt]
\textbf{System:} You are an audio quality verifier. Given a candidate
audio clip and its target sub-event, decide whether the clip can be
inserted into the surrounding scenario without breaking acoustic
consistency. The same prompt is applied identically to retrieved clips
and to clips synthesized by AudioX or ElevenLabs --- verification must
be source-agnostic so the two paths cannot drift apart over time. Be
conservative: when unsure, prefer \texttt{reprocess} over
\texttt{accept}, and \texttt{reject} over \texttt{reprocess}.\\[2pt]
\textbf{Decision criteria:}
\begin{itemize}
  \item \textbf{Identity}: the dominant sound source is the target
        sub-event itself, not a co-occurring background. A clip
        captioned ``door slam'' but dominated by background music fails.
  \item \textbf{Cleanliness}: no overlapping speech, music, or alarm
        sounds that would conflict with the named adjacent sub-events;
        ambient noise from real recordings is acceptable.
  \item \textbf{Duration fit}: clip length within
        $[0.5\times,\ 2\times]$ of \texttt{rough\_duration\_s}; if too
        long, label \texttt{reprocess} so TFJP can localize the
        informative span.
  \item \textbf{Continuity}: onset and offset are not abruptly clipped
        mid-event; if so, label \texttt{reprocess} (TFJP can repair the
        boundary) rather than \texttt{reject}.
\end{itemize}
\textbf{User:} \texttt{scenario}, \texttt{target\_sub\_event},
\texttt{adjacent}=\verb|({prev_event},{next_event})|,
\texttt{clip\_caption} (auto-generated),
\texttt{clip\_source}$\in$\{\verb|retrieval|,\verb|AudioX|,\verb|ElevenLabs|\}.\\[2pt]
\textbf{Required output (JSON):}
\begin{itemize}
  \item \texttt{decision}: \texttt{accept} $|$ \texttt{reprocess} $|$ \texttt{reject}.
  \item \texttt{reason}: one short sentence explaining the decision in
        terms of the four criteria above.
  \item \texttt{conflicts}: list of adjacent sub-event names this clip
        would conflict with if accepted (empty list if none).
\end{itemize}
\textbf{Cascade:} \texttt{accept}$\to$commit to scenario;
\texttt{reprocess}$\to$re-route through TFJP, then re-verify once
(at most one reprocess per clip); \texttt{reject}$\to$advance to next
retrieval candidate, falling back to generation via
\texttt{fallback\_caption} after all three retrieval candidates exhaust.
\end{tcolorbox}
\caption{Prompt template for hierarchical event curation, Part 2: clip
grounding verification, applied identically to retrieved and synthesized
clips so the two paths share one acceptance criterion.}
\label{prompt:curation-p2}
\end{figure}


\begin{figure}[h]
\centering
\begin{tcolorbox}[
  colback=white, colbacktitle=white, colframe=black,
  coltitle=black,
  title=\textbf{Prompt Template (Comprehension-Aware Supervision)},
  fonttitle=\bfseries,
  boxrule=0.5pt, arc=2pt,
  fontupper=\small,
  left=8pt, right=8pt, top=4pt, bottom=4pt
]
\textbf{Prompt A --- History-Review Question Generation}\\[1pt]
\textbf{System:} Write one natural follow-up question whose answer
depends on information from at least three rounds earlier in the
dialogue, and cannot be inferred from common knowledge or the latest
1--2 turns alone.\\[2pt]
\textbf{Constraints:}
\begin{itemize}
  \item Use an answer-supplying turn at least 3 rounds back.
  \item Avoid scaffolding cues such as ``as you mentioned earlier''.
  \item Prefer entity-grounded questions over open-ended ones.
\end{itemize}
\textbf{User:} \texttt{dialogue\_history}: \verb|{dialogue_history}|.\\[2pt]
\textbf{Required output (JSON):} \texttt{question}, \texttt{reference\_answer},
\texttt{reference\_turn\_idx} (integer index of the answer-supplying turn).

\rule{\linewidth}{0.4pt}\\[2pt]

\textbf{Prompt B --- Silent-Audio Verification}\\[1pt]
\textbf{System:} Decide whether a streaming audio assistant should
respond or remain silent. Default to \texttt{silent}; use
\texttt{respond} only for high-stakes events: \textbf{(i)} acute
physiological distress, \textbf{(ii)} severe weather or natural hazard,
\textbf{(iii)} equipment damage or fire indicator, or \textbf{(iv)}
safety-critical environmental signal. Use \texttt{borderline} for
genuinely ambiguous clips.\\[2pt]
\textbf{User:} \texttt{scenario}: \verb|{scenario}|;\quad
\texttt{caption}: \verb|{caption}|;\quad
\texttt{adjacent}=\verb|({prev_event},{next_event})|.\\[2pt]
\textbf{Required output (JSON):} \texttt{label},
\texttt{triggered\_criterion}, and \texttt{reason}.
\end{tcolorbox}
\caption{Prompt template for comprehension-aware supervision:
history-review question generation (Prompt A) and silent-audio
verification (Prompt B). Both run on the same chat LLM in JSON-mode
decoding; \texttt{borderline} clips from Prompt B are discarded rather
than mislabeled.}
\label{prompt:supervision}
\end{figure}

\begin{figure}[h]
\centering
\begin{tcolorbox}[
  colback=white, colbacktitle=white, colframe=black,
  coltitle=black,
  title=\textbf{Prompt Template (Spoken-Style Rewriting)},
  fonttitle=\bfseries,
  boxrule=0.5pt, arc=2pt,
  fontupper=\small,
  left=8pt, right=8pt, top=4pt, bottom=4pt
]
\textbf{System:} You are a script rewriter preparing text for
text-to-speech rendering. Convert the input into a form that, when
spoken aloud by CosyVoice, sounds natural, fluent, and free of
visually-formatted artefacts. The output is fed directly to the TTS
engine; downstream ASR will then verify that the rendered audio is
recoverable to within a WER threshold of $\tau_{\text{wer}}\!=\!0.10$
against your rewrite. Therefore: the rewrite must be unambiguously
pronounceable, and meaning-bearing words must survive the round-trip.\\[2pt]
\textbf{Rewriting rules:}
\begin{itemize}
  \item \textbf{Markdown strip}: remove all markdown syntax, code
        fences, bullet symbols, URLs, file paths, hashtags, and emojis;
        replace inline code spans with their plain-text content.
  \item \textbf{Numeral expansion}: ``$100$'' $\to$ ``one hundred'';
        ``$3.14$'' $\to$ ``three point one four''; phone numbers and
        years remain digit-by-digit (``2025'' $\to$ ``two thousand
        twenty-five'' or ``twenty twenty-five'' depending on register).
  \item \textbf{Abbreviation expansion}: ``Dr.''$\to$``Doctor'';
        ``Inc.''$\to$``Incorporated''; ``etc.''$\to$``and so on'';
        Latin abbreviations (``e.g.'', ``i.e.'') are spelled out as
        ``for example'' / ``that is''.
  \item \textbf{Symbol replacement}: ``\&''$\to$``and'';
        ``\%''$\to$``percent''; ``\$''$\to$``dollars''; ``\#''$\to$
        ``number'' or ``hash'' depending on context; mathematical
        operators are read aloud (``$+$''$\to$``plus'').
  \item \textbf{Length and meaning}: preserve original meaning, tone,
        and approximate length; do not summarize or paraphrase
        content-bearing words; if the input contains code blocks longer
        than three lines, replace the whole block with ``code omitted''
        rather than reading it literally.
  \item \textbf{No-op pass-through}: if the input is already a clean
        spoken sentence, return it unchanged with empty
        \texttt{changes}.
\end{itemize}
\textbf{User:} \texttt{text}: \verb|{text}|.\\[2pt]
\textbf{Required output (JSON):}
\begin{itemize}
  \item \texttt{rewritten}: final spoken-form text.
  \item \texttt{changes}: list drawn from \{\texttt{markdown\_strip},
        \texttt{numeral\_expand}, \texttt{abbrev\_expand},
        \texttt{symbol\_replace}, \texttt{punctuation\_normalize},
        \texttt{code\_block\_drop}\}; empty list if unchanged.
\end{itemize}
\end{tcolorbox}
\caption{Prompt template for the spoken-style rewriter applied to
text-form supervision sources (MOSS, GammaCorpus, instruction chats)
prior to CosyVoice rendering. The WER round-trip via downstream ASR
constrains how aggressively the rewriter may paraphrase.}
\label{prompt:tts}
\end{figure}

\end{document}